\documentclass[reprint,english,superscriptaddress,secnumarabic,amssymb,amsmath,nobibnotes,aps,prd,showkeys,showpacs,longbibliography,floatfix]{revtex4-1}

\usepackage{xcolor}
\usepackage{enumerate}
\usepackage{graphicx}
\usepackage{overpic}
\usepackage{placeins}
\usepackage{verbatim}

\begin{document}

\title{Disk sources of the Kerr and Tomimatsu-Sato spacetimes: construction and physical properties}

\author{Tom\'a\v{s} Ledvinka}
\email{tomas.ledvinka@mff.cuni.cz}
\affiliation{Institute of Theoretical Physics, Faculty of Mathematics and Physics, Charles University in Prague, V Hole\v{s}ovi\v{c}k\'ach 2, 180 00 Praha 8, Czech Republic}

\author{Ji\v{r}\'i Bi\v{c}\'ak}
\email{jiri.bicak@mff.cuni.cz}
\affiliation{Institute of Theoretical Physics, Faculty of Mathematics and Physics, Charles University in Prague, V Hole\v{s}ovi\v{c}k\'ach 2, 180 00 Praha 8, Czech Republic}
\affiliation{Max Planck Institute for Gravitational Physics (Albert Einstein Institute), Am M\"uhlenberg 1, Potsdam 14476, Germany}

\begin{abstract}
We construct the disk sources matched to the exact vacuum Kerr and to the two classes of Tomimatsu-Sato spacetimes. We analyze two models 
of the matter forming these disks. At each radius we consider either a rotating massive ring with pressure or two counter-rotating streams of particles in circular geodesic motion. Dragging effects present in such
spacetimes lead either to rotation of rings or asymmetry of both streams. We demonstrate that the model of rotating rings is general enough to describe all axisymmetric stationary disk sources with vanishing radial pressure which satisfy weak energy condition, 
and that centrifugal effects present in the disk sources of spacetimes with large angular momentum prevent
the construction of highly compact sources made of counter-rotating streams of geodesic particles. We illustrate
the radial distribution of the mass inside the disks and the angular velocities of both geodesic streams.
\end{abstract}

\maketitle

\section{Introduction}
Gravitating disks and their fields are of great intrinsic interest and have important astrophysical 
applications. In Newtonian theory one can easily build potential-density pairs for axisymmetric disks by using 
the method of images common in the galactic dynamics. A point mass placed at a distance $b$ below the centre
$\rho = 0$ of a plane $z = 0$ gives a solution of Laplace’s equation above the plane. Then, considering
the potential obtained by reflecting this $z \geq 0$ potential in $z = 0$, a symmetrical
solution both above and below the plane is obtained which is continuous but has a discontinuous normal derivative on 
$z = 0$. The jump gives a positive surface density on the plane. In galactic dynamics one considers
general line distributions of mass along the negative $z-$axis and, employing
the device described above, one finds the potential-density pairs for general
axially symmetric disks. In case of static, axisymmetric vacuum spacetimes in
general relativity, as Hermann Weyl has shown in 1917 already, exact solutions of the Einstein equations can, 
in suitable (cylindrical-type) coordinates, be generated from vacuum solutions of the Laplace equation; hence, 
classical potentials can be used to generate relativistic solutions. In \cite{BicakLyndPichon93}, 
an infinite number of new static solutions
of the Einstein equations were found starting from realistic potentials used to
describe flat galaxies, as given by Evans and de Zeeuw \cite{Evans}.
Although these disks are Newtonian at large distances, in their central
regions interesting relativistic features arise, such as velocities close to the
velocity of light, and large redshifts. In a more mathematical context, a number of exact static axisymmetric 
(electro)vacuum solutions of the Einstein equations which involve singularities in their central regions can be interpreted as exact external fields of disk sources. In \cite{BicakLyndKatz93} it is shown that most vacuum static Weyl solutions, including the Curzon and the Darmois-Vorhees-Zipoy 
solutions can arise as the metrics of counter-rotating relativistic disks. In order that the disks may be seen as being produced by circulating particles one has to consider them as being made of two equal collisionless streams that circulate in opposite directions around the center. In this way one avoids the effect of dragging of inertial frames which is absent in the static metrics.

Because the dragging of inertial frames implies non-linear fields outside the disks it is very hard to find analytically expressed exact stationary, asymptotically flat solutions of the Einstein equations corresponding to rotating material sources. Remarkably, the Riemann-Hilbert problem (the boundary value problem) for an infinitesimally thin, finite disk of dust particles which rotate rigidly around a common center was formulated and solved (see \cite{Neugebauer1995,Neugebauer1996}  and references therein). Later it was generalized to the solutions representing counter-rotating streams \cite {Klein1999,Klein2001} which in one limit go over to the rigidly rotating disk of the preceding case, in the other limit--to the static counter-rotating finite disks described in \cite {MorganMorgan}. Although these solutions are very valuable, they are special, and it is so worthy to construct other type of rotating disk sources.   

The method of images cannot be used directly because of non-linear fields outside the disks due to dragging. However, the method of images described above may be viewed as the  \textit{identification} of the surface $z=b$ with the surface $z=-b$. The field remains continuous but the jump of its normal derivatives induces a matter distribution in the disk. We shall describe and illustrate the procedure in detail in the following section and apply it throughout the paper. Hence, if a stationary vacuum solution containing some pathological regions in the central parts is available, one can always ``cut the regions off'' and end-up with a disk source surrounded by a regular spacetime. It is thus worth using much wider family of vacuum solutions with `sources' in the form of various singularities or other pathologies to obtain spacetimes with idealized but much more realistic sources in the form of shells and disks. In this way one can substantially enlarge the set of spacetimes on which the complicated  relation between 
source and spacetime parameters can be studied. There is quite extended literature on disk sources constructed by this type of method in vacuum spacetimes (see, e.g., \cite{BLPRL93,Pichon96,VogtLet1,Gonzales2012}), in electrovacuum spacetimes (e.g. \cite{Antonio,Zofka}, for a review, see \cite{Klein2003}), in spacetimes representing a disk with a black hole inside (see, e.g. \cite{Lemos,VogtLet2}). Those are just casually selected references, other citations can be found therein. For a comprehensive review on self-gravitating relativistic disks, in particular those around 
black holes, see \cite{KarasHure} and more recent work by Semer\'ak, Sukov\'a and collaborators on 
disks and rings around black holes (for example, the superpositions of a Schwarzschild black hole 
with the inverted Morgan-Morgan counter-rotating thin rotating thin-discs)--see \cite{Semerak}
and references therein.

In this paper we consider disks without radial pressure as sources of the first three ($\delta=1,2,3$) members of the 
Tomimatsu-Sato (TS) class of spacetimes. These are asymptotically flat stationary solution of Ernst equation given by rational functions in spheroidal coordinates \cite{TomimatsuSato,TomimatsuSatoNC73}. The TS $\delta=1$ metric is the well known Kerr solution of Einstein's equations --  
the only asymptotically flat stationary spacetime which, under suitable choice of parameters, represents a rotating black hole with all singularities hidden under the event horizon. Very soon after the discovery of the TS solutions, their properties (for $\delta=2$) were investigated in \cite{Gibbons}. It was shown that the solutions represent a spinning naked singularity with causality violating regions and for parameter $q=0$ (see below) it reduces to the Vorhees solution. 
The complicated structure of TS $\delta=2$ spacetime centre led to the general belief that there is no horizon present there, but its location and shape was revealed in
2003 when Kodama and Hikida  investigated in depth the global structure of both Zipoy-Vorhees-Weyl and the $\delta=2$ TS spacetimes in \cite{Kodama}. 

Propertites of the central strong-field regions of TS spacetimes are also investigated in this paper where we pose questions such as:
How strong gravitational field can be created without violating energy conditions, how large total angular momentum can a disk producing a TS solution posses, or what prevents us to construct disk sources which reveal closed timelike curves regions known to be present in  the central parts of these spacetimes.
To summarize our main motivations, we are focused on questions of principle, as 
(i) whether there are physically plausible matter sources of known vacuum solutions, often containing naked
 singularities like in the TS case;
(ii) what effects and possible pathologies arise due to very 
strong gravitational fields described by exact, explicit, hence quite idealized,
general-relativistic models of disks formed by circular counter-rotating streams or rotating rings with tangential pressure.

Although the work on counter-rotating relativistic disks is far from the realistic models of galaxies, the counter-rotating disks
have become more attractive from the beginning of 1990's when counter-rotating disk galaxies were first observed
\cite{Rubin}. Since then the counter-rotating components were detected in ``tens of galaxies along all the Hubble sequence, from 
elliptical to irregulars'' (see \cite{Corsini} for a review from 2014). For  example, among 53 lenticular (S0) galaxies 17 
had counter-rotating gas, i.e. 32\%, less than 10\% host a significant fraction of counter-rotating stars.
In the work \cite{Bass} from 2017 on the formation of S0 galaxies, the authors cite that 
``the observation of counter-rotation in galaxies is becoming more commonplace...the percentage of S0 galaxies
that exhibit counter-rotation is 20-40 per cent''. Mergers appear as the prime candidate for the origin of counter-rotation, however, accretion and other 
mechanisms are also proposed to explain the formation of such systems. Moreover, recently 
counter-rotating disks arising from the accretion of counter-rotating gas on the surface of a co-rrotating disk have been considered \cite {Dyda}. The authors give several scanarios how such situations may arise on a stellar mass scale. On a larger scale they refer to the King and Pringle \cite{KiPr} idea in which the rapid growth of the massive black holes at high redshifts will be compatible with the Eddington limit of their emission if the gas accretion will proceed in sequence of clouds with varying angular momenta. 
    
In the following section we show how the method of constructing the disk sources by performing identifications of spacetime hypersurfaces works. The metric induced on the identification hypersurface is invariantly expressed in terms of an appropriate tetrad. The tetrad components of the stress-energy tensor of the disk matter are given in terms of the jumps of the extrinsic curvature of the hypersurface and the discontinuities of metric functions (gravitational potentials). The masses and the angular momenta of the disks are expressed using the Komar integrals. By the procedure described one can always find a disk source for a given solution of the Einstein equations. Often, however, the source will have not be realizable by a physically meaningful matter. In Section III we thus formulate criteria (like ``energy conditions'') guaranteeing a physically meaningful source. Physically clear and most plausible sources are those made from counter-rotating streams of incoherent dust moving along circular geodesics. This model is discussed in detail in Section IV. 
In Section V we summarize main properties of Tomimatsu-Sato spacetimes.
In the most extended Section VI, the properties of the disks are analyzed 
by both analytical and numerical methods; and they are illustrated by a number of figures. The section contains four subsections, one on general Tomimatsu-Sato solutions, the following on the Kerr solutions (TS solutions with index $\delta=1$), with a separate discussion of the disks producing extreme Kerr metrics; the last two subsections describe and illustrate the properties of the TS disks producing the TS metrics with $\delta=2$ and $\delta=3$. The main text comes to an end by conclusions in which main results of the paper are summarized. 
Two Appendices, containing the discussion of the static disks and the derivation of the total mass and total angular momentum of the disks from the Komar integral, are added finally.

\section{Stationary axisymmetric spacetime with disk sources}
 
\begin{figure}[!htb]
\centerline{
\includegraphics[width=8.5cm]{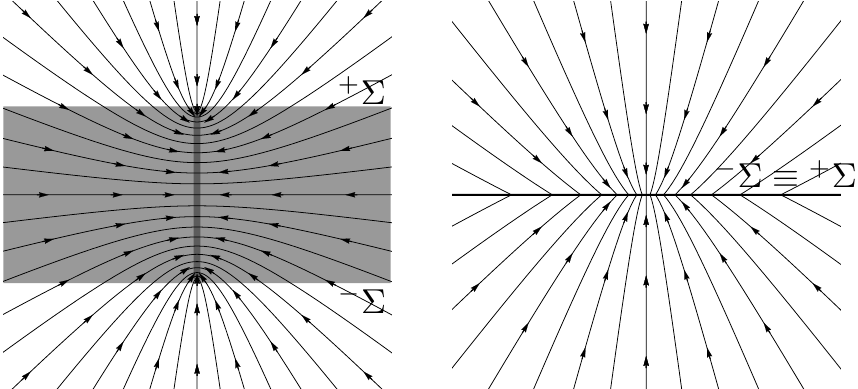}
} 
\caption{When the central region between ${}^-\Sigma$ and ${}^+\Sigma$  
(left, shaded) containing an arbitrary source
of the field  is removed, we obtain the field of a disk source (right).
\label{FIG1}}
\end{figure}

We start from the metric given in Weyl's canonical form in the Weyl-Papapetrou (WP) coordinates \{$t,\phi,\rho,z$\},
 see e.g. \cite{Stephani2009},
\begin{equation}
 d{s}^2 =  - e^{2\nu} (dt+A d\phi)^2 +  \frac{e^{2\zeta}}{e^{2\nu}}(d\rho^2+dz^2) +\frac{\rho^2}{{e^{2\nu}}} d\phi^2
 \label{WeyPap},
\end{equation}
where functions $\nu,\zeta, A$ depend on $\rho,z$ only. We assume the metric to 
represent stationary axisymmetric asymptotically flat vacuum solution of the Einstein equations
`above' the axisymmetric hypersurface ${}^+\Sigma$ parametrically 
described by
\begin{equation}
 z=z(s), ~~ \rho=\rho(s)\label{zrho_s},~~~ \phi,~~~ t,
\end{equation}
where $s$ is a `radial' coordinate from the center of the disk for any $t=\rm const$. The spacetime `below' the surface is then completed 
assuming the symmetry of all potentials under the reflection  
\begin{equation}
 t\rightarrow t, ~~ \rho\rightarrow  \rho,~~ \phi\rightarrow \phi, ~~z\rightarrow -z. \label{identrule}
\end{equation}
This completion also yields hypersurface ${}^-\Sigma$. For spacetimes globally satisfying the symmetry
 \eqref{identrule}, this construction is equivalent to the 
exclusion of the `central' region between ${}^-\Sigma$ and ${}^+\Sigma$; see Fig. \ref{FIG1} illustrating
the identification (coordinates $t$ and $\phi$ being suppressed).

The hypersurface ${}^+\Sigma$ has unit normals
given by the following spacetime components
\begin{align}
 n^\mu = \lambda(s) \left(\rho'(s)\delta_{z}^\mu-z'(s) \delta_{\rho}^\mu\right),
\\ 
  \lambda(s) = \frac{e^{\nu-\zeta}}{\sqrt{{\rho'(s)}^2+{z'(s)}^2}}.
\end{align}
We use the following convention for indices: $k,l,m,...,\kappa,\lambda,\mu,... = 0,..,3$, $a,b,c,...,
\alpha,\beta,\gamma,... = 0,1,2$, $A,B,... = 0,1$.
If the coordinates $x^\mu=\{t,\rho,\phi,z\}$ are used as indices they always mean the particular component  ($x^3\equiv x^z= z$),
or the partial derivative with respect to a particular coordinate ($g_{ab,z}=\partial g_{ab}/\partial z$). 
Tetrad (projected) components of tensors use brackets (see, e.g. \eqref{g_ab}).
Induced metric on the 3-dimensional hypersurface ${}^+\Sigma$ is simply given by metric \eqref{WeyPap} with
the substitution $e^{2\zeta-2\nu}(d\rho^2+dz^2) \rightarrow ds^2/\lambda^2$.
If $s$ is the proper radius $\lambda=1$.

Following the approach to singular hypersurfaces and thin shells initiated by W. Israel, \cite{Israel66},
\cite{Barrab}, see also \cite{Kuchar68} for in general charged shells,
we define an orthonormal basis tangent to ${}^+\Sigma$    
\begin{align}
e^\mu_{(t)} = e^{-\nu}\delta^\mu_{t},~~~
e^\mu_{(\phi)} = \frac{e^{\nu}}{\rho} \left(\delta^\mu_{\phi}-A\delta^\mu_{t}\right),\\
e^\mu_{(\rho)} = \lambda(s) \left(\rho'(s)\delta_{\rho}^\mu+z'(s) \delta_{z}^\mu\right),
\label{basis123}
\end{align}
$e^\mu_{(a)} n_\mu =0$, so that $n_\mu$ can used as the tetrad vector $e^\mu_{(z)}$ associated with the $z$-coordinate.
The tetrad components of the induced metric are simply
\begin{equation}
\gamma_{(a)(b)} = g_{\mu\nu} e^\mu_{(a)} e^\nu_{(b)}={\rm diag}(-1,1,1)
\label{g_ab}
\end{equation}
($a,b = \{t,\rho,\phi\}$). 
The extrinsic curvature is
given by the derivative of the normal field along $\Sigma={}^+\Sigma \equiv {}^-\Sigma$:
\begin{equation}
K_{(a)(b)} =  e^\mu_{(a)} e^\nu_{(b)}\nabla_\mu n_\nu
           =  - n_\nu e^\mu_{(a)} \nabla_\mu e^\nu_{(b)}.
\label{KabDEF}           
\end{equation}
The metric \eqref{g_ab} is used 
to lower/raise indices for 3-tensor quantities, e.g.,
$K^{~(c)}_{(c)} = K_{(a)(b)} \gamma^{(a)(b)}$.

The surface stress-energy tensor which arises
due to the identification along the hypersurfaces ${}^\pm \Sigma$ with identical 
induced metric ${}^+g_{(a)(b)}={}^-g_{(a)(b)}$ is given 
by the difference of the extrinsic curvatures of the hypersurfaces 
$\left[ K_{(a)(b)} \right] = {}^- K_{(a)(b)} - {}^+ K_{(a)(b)}$
by the relation \cite{Israel66,Barrab,Kuchar68} 
\begin{equation}
S_{(a)(b)} = (8\pi)^{-1} \left( [K_{(a)(b)}] -  \gamma_{(a)(b)} [K^{~(c)}_{(c)}] \right).
\end{equation}

Performing the identification
\eqref{identrule}, the opposite normals have opposite signs and 
we get $\left[ K_{(a)(b)} \right] = -2\;{}^+K_{(a)(b)}$.

Even though the static tetrad \eqref{basis123} does not exist inside the ergoregions it  yields the clearest formulas for 
the surface stress-energy tensor which arises after the identification \eqref{identrule}: 
\begin{align}
 4\pi S_{(t)(t)} =& ~2\nu_{,n} - \zeta_{,n}  + (\kappa_1 + \kappa_2)e^{\nu-\zeta},
 \label{tetStt}
 \\
 4\pi S_{(\phi)(\phi)} =& ~\zeta_{,n} - \kappa_2e^{\nu-\zeta},
\\
8\pi S_{(t)(\phi)}=&~{\rho}^{-1} e^{2\nu} A_{,n} ,
\label{tetStf}
\\
4\pi S_{(\rho)(\rho)}=&-  \kappa_1 e^{\nu-\zeta}\label{Srhorho},
\end{align}
where $f_{,n} =n^\mu \partial_\mu f = \lambda (\rho' f_{,z}-z' f_{,\rho})$ denotes normal derivative on $\Sigma^+$ and 
\begin{equation}
\kappa_1 = \frac{1}{\rho} \frac{z'}{\sqrt{{\rho'}^2+{z'}^2}},~~~
\kappa_2 = \frac{\rho'z''-z'\rho''}{({\rho'}^2+{z'}^2)^{\frac32}} 
\end{equation}
are the principal curvatures of the surface given by $t=\rm const.$ in the hypersurface ${}^+\Sigma$ when considered in flat Euclidean space. Thus, the surface stress-energy tensor is given by a combination 
of the discontinuities in gravitational potentials and the geometric properties of $\Sigma$.
An important exception is the radial stress $S_{(\rho)(\rho)}$ which is up to a factor $e^{\nu-\zeta}$ equal to the value of  $S^{\rm flat}_{(\rho)(\rho)}$ in the flat spacetime with $\nu=\zeta=A=0$. In this sense radial stresses here reflect rather the external curvature of $\Sigma$ in flat space than the gravitation of sources inside the cut-off region between ${}^\pm \Sigma$. (Notice that the formulae equivalent to Eqs. \eqref{tetStt}-\eqref{Srhorho} but given in a coordinate basis, as in \cite{Pichon96}, do not reflect so clearly the respective influences of gravitational field and of the identification surface.)

Later in the following  we wish to study in detail the properties of the disks without radial pressure. These enable the 
clearest interpretation of the surface stress-energy tensor.
Due to the properties of the WP coordinates which imply Eq. \eqref{Srhorho} such disks correspond to 
the hypersurfaces given by $z(s) = b = \rm const$.
Then the surface stress-energy tensor $S_{ab}$, given in the coordinate rather than tetrad components,
simplifies to 
\begin{equation} \label{Sabzkonst}
{S}_{\alpha \beta} = -~{\sqrt{g_{\rho\rho}}\over 8 \pi} 
\left( {{g}_{\alpha \beta}\over g_{\rho\rho}} \right)_{\!\!,z}.
\end{equation}
This is the form, which holds only in WP coordines, and unifies Eqs. \eqref{tetStt}-\eqref{Srhorho}. Since in WP coodrinates $g_{\rho\rho}=g_{zz}$, indices $\alpha,\beta=t,\rho,\phi$ can be replaced with $\mu,\nu=t,\rho,z,\phi$ and we get a tensor tangential to the disk hypersurface. Once $S_{\mu\nu}$ is computed in the WP coordinates, we can simply obtain its components in other coordinates or an arbitrary tetrad by appropriate transformations or projections.  
However, inside the ergoregions we have either to work with coordinate components \eqref{Sabzkonst}, or  use the well-known ``zero-angular-momentum'' (ZAMO) tetrad
(see, e.g., \cite{Wald84}) rather than  \eqref{basis123} which becomes unphysical there:
\begin{align}
\tilde e^\mu_{(t)} &= \frac{g_{\phi\phi}^{1/2}}{\rho} \delta^\mu_{t} - \frac{g_{t\phi}}{\rho g_{\phi\phi}^{1/2}} \delta^\mu_{\phi} ,
\nonumber
\\
\tilde e^\mu_{(\phi)} &= \frac{1}{g_{\phi\phi}^{1/2}} \delta^\mu_{\phi},~~~
\tilde e^\mu_{(\rho)} = \frac{1}{g_{\rho\rho}^{1/2}} \delta^\mu_{\rho},~~~
\tilde e^\mu_{(z)} = n^\mu.
\label{basisZAMO}
\end{align}

The method used for the construction of our disk sources guarantees
that the values of the total mass $M_D$ and angular momentum $J_D$ of the disks (given by the asymptotic
behavior of the metric) are equal to those of the original
spacetimes, $M$ and $J$, provided that the metric satisfies the vacuum Einstein equations outside the disk.
Using Komar integrals \cite{Wald84} in stationary axisymmetric spacetimes we can write these quantities
as integrals over certain surface densities over the disks (see also Appendix B):
\begin{align} 
\label{KomarMDX} M_D &= {1\over 4 \pi}\int_D \left( 2\nu_{,n} + 
\rho^{-2} e^{4\nu}A A_{,n}
\right) ~\rho\; \lambda^{-1}(s) \;ds \;d\phi,
\\
\label{KomarJDX} J_D &=-{1\over 8 \pi}\int_D\big[
4A\, \nu_{,n}  +A_{,n}
\left(1+
\rho^{-2} A^2 e^{4\nu}
\right) \nonumber\\
&~~~~+  2A \kappa_1 e^{\nu-\zeta}  
\big]  \;\rho\; \lambda^{-1}(s)\; ds\;  d\phi.
\end{align}
For all spacetimes considered we found both integrands regular also inside the ergoregions where
$e^{2\nu}<0$ and individual terms, such as $\nu_{,n} = (e^{2\nu})_{,n}/(2e^{2\nu})$, diverge at circles
where the disk source enters the ergoregion.

\section{Energy conditions and the interpretation of the surface
stress-energy tensor of the disk}
For a given solution of the Einstein equations the method we use always provides \textit{some} source. 
In many cases, however, its properties will be unphysical, for example, it will have a  negative energy
density or it will move with superluminal speeds. 
One way to guarantee physically acceptable sources is to demand that they satisfy some type of suitable
energy conditions introduced originally in the studies of the global properties of spacetimes
(see, e.g., \cite{Wald84}). They can be employed without further assumptions about the `material' creating the surface stress-energy tensor $S_{\mu\nu}$. The weak energy condition (WEC) requires, that any observer 
with her velocity $W^\mu$ must observe a non-negative energy density $S_{\mu\nu} W^\mu W^\nu$. The dominant energy condition (DEC) is based on the properties of the energy-momentum current $- S_{\mu\nu} W^\mu$, which should be future directed timelike vector for classical matter; in fact, DEC $\implies$ WEC. 

It is much easier to decide whether a certain energy condition holds 
if the stress-energy tensor is in diagonal form.
The diagonalization may be achieved either by finding an observer
who measures the surface stress energy tensor to be diagonal, or by solving 
the following eigenvalue problem
\begin{equation} \label{DefDiag}
 \left(S_{(a)(b)} - \lambda \eta_{(a)(b)}\right) X^{(b)} = 0.
 \end{equation} 
Here we assume an arbitrary orthonormal 3-basis satisfying $e^\mu_{(a)} e_{\mu(b)} = \eta_{(a)(b)} $
and the stress-energy tensor with only one off-diagonal component, $S_{(t)(\phi)}$.
Then the characteristic equation has real solutions if ($A,B=t,\phi$)
\begin{equation} \label{DefDeltaWEC}
 \sigma  = \left(S_{(t)(t)} - S_{(\phi)(\phi)}\right)^2 + 4\;{\rm det}~S_{(A)(B)} \ge 0.
\end{equation} 
This inequality is also the necessary condition for the stress-energy tensor to satisfy the 
weak energy condition, because ${\rm WEC}\implies \sigma\ge 0$. [Since $\sigma = 
(S_{(t)(t)}-2S_{(t)(\phi)}+S_{(\phi)(\phi)}) (S_{(t)(t)}+2S_{(t)(\phi)}+S_{(\phi)(\phi)}) =
(S_{\mu\nu}\xi_1^\mu \xi_1^\nu)(S_{\rho\sigma}\xi_2^\rho \xi_2^\sigma)$ for certain two null vectors $\xi_{1,2}^\mu$, 
then, if $\sigma<0$, one of the two projections of $S_{\mu\nu}$ is negative and, from continuity, there will be observers with velocity near either $\xi_{1}^\mu$ or $\xi_{2}^\mu$ seeing negative energy density.]
From \eqref{DefDiag} we obtain three real eigenvalues 
$-\mu$, $p_{\phi'}$ and $p_r$       where
\begin{align}
\mu=& \frac{1}{2}\left( S_{(t)(t)} - S_{(\phi)(\phi)} + \sqrt\sigma \right),\\
p_{\phi'} =&  \frac{1}{2}\left( S_{(\phi)(\phi)} - S_{(t)(t)} + \sqrt\sigma \right),\
p_r=S_{(\rho)(\rho)}.
\end{align}
The diagonalization yields the tetrad (frame) vectors of the observer with respect to whom the 
stress-energy tensor is in the diagonal form, 
\begin{align}
e_{(t')}^\mu &=& \left(e_{(t)}^\mu+v\; e_{(\phi)}^\mu\right)/{\sqrt{1-v^2}},~~~~~
e_{(\rho')}^\mu&=e_{(\rho)}^\mu,
\nonumber\\
e_{(\phi')}^\mu &=& \left(v\;e_{(t)}^\mu+e_{(\phi)}^\mu\right)/{\sqrt{1-v^2}},~~~~
e_{(z')}^\mu &=e_{(z)}^\mu,
\label{FIOtetrad}
\end{align}
where  $v = -2 S_{(t)(\phi)}/\left({S_{(t)(t)}+S_{(\phi)(\phi)}+\sqrt{\sigma}}\right)$,
normalized so that $e^\mu_{(\alpha')} e_{\mu(\beta')} = \eta_{(\alpha')(\beta')}$.
Then 
\begin{equation} \label{DefDiagmup}
   S^{\mu\nu} = \mu ~e_{(t')}^\mu e_{(t')}^\nu + p_{\phi'}~e_{(\phi')}^\mu e_{(\phi')}^\nu + p_{r}~ e_{(\rho)}^\mu e_{(\rho)}^\nu~.
\end{equation} 
If the decomposition (\ref{DefDiagmup}) of $S_{\mu\nu}$ is achievable, 
we can examine $\mu,p_r, p_{\phi'}$ and decide whether energy conditions are satisfied. The the weak energy condition reads
\begin{equation} \label{wecdiag}
 \mu \ge 0~\wedge~p_r \ge -\mu~\wedge~p_{\phi'} \ge -\mu,
\end{equation} 
and the dominant energy condition is equivalent to
\begin{equation} \label{decdiag}
 \mu \ge 0~\wedge~p_r^2+p_{\phi'}^2  \le \mu^2.
\end{equation} 
As we also have relations
\begin{equation} 
{\rm WEC} \implies S_{(t)(t)} + S_{(\phi)(\phi)} \ge 0 \wedge {\sigma} \ge 0 \implies |v|<1,
\label{ECimplicatons}
\end{equation} 
we see that if ${\sigma}<0$ or $|v|>1$ no one energy condition is satisfied.

Instead of the velocity $v$ with respect to an observer equipped with some tetrad frame,
we can, as we did in \cite{BLPRL93}, consider the angular velocity $\Omega=e_{(t')}^\phi/e_{(t')}^t$
with respect to a static observer at infinity. Regarding the properties of tetrad basis from which 
one gets $\sigma = (S_A^{~A})^2-4\;{\rm det~}S_A^{~B}$,
we obtain
\begin{align}
\label{Omega_FIO}
\Omega = -\frac{2 S_t^{~\phi}}
{S_\phi^{~\phi}-S_t^{~t}+\sqrt{\sigma}}.
\end{align}
In weak gravitational fields this angular velocity gives eigenvector $e_{(t')}^\mu$ of \eqref{DefDiag} and the one with opposite sign in front of $\sqrt{\sigma}$ provides a ratio of $\phi$ and $t$ components of $e_{(\phi')}^\mu$.
This may change in the central regions of strong-field disks where one has to check which of the eigenvectors is timelike. 

For vanishing radial stresses, $p_r=0$, the diagonalization described above also yields a possible 
model of the surface stress-energy tensor of the disks.
The equatorial plane can be regarded as being divided into massive circular rings, 
each with the four-velocity $e^\mu_{(t')}$ and the surface energy density $\mu$. These rotating rings (RR) are supported against the collapse or expansion by their internal azimuthal surface stress $p_{\phi'}$.
Although usual materials do not provide $|p|/\mu \lesssim 1$, such high stresses are consistent with 
energy conditions. In addition, a model more general than RR is not needed -- when its diagonalisation is not possible and the surface stress-energy tensor cannot be made of RR, since, due to \eqref{ECimplicatons}, neither DEC nor WEC holds.

\section{Disks from counter-rotating surface streams of geodesic particles}

One of the simplest matter models in general relativity is incoherent dust.
To interpret disk sources of {\it static} spacetimes, Morgan and Morgan \cite{MorganMorgan} introduced 
the simple model of two counter-rotating geodesic streams (CRGS) of dust in stationary circular motion. 
In these static spacetimes no dragging is present and the surface stress-energy tensor is diagonal 
with two nonzero components, $S^{tt}$ and $S^{\phi\phi}$; 
the latter being generated by the velocity of counter-rotating particles.
The fact that the velocity producing $S^{\phi\phi}$ component matches the Keplerian velocity
and the assumed circular motion of counter-rotating particles is consistent with geodesic motion can, as in \cite{MorganMorgan}, be viewed as an implication of the restriction 
$ (S^{t}_{~t} - S^{\phi}_{~\phi} ) \rho \nu_{,\rho} = -S^{\phi}_{~\phi}$ 
which Einstein's equations impose on the stress-energy tensor.
(The construction of the disk sources of static axisymmetric spacetimes is briefly explained in Appendix A.)

We wish to discuss how the concept of disk surface streams made of counter-rotating particles 
on circular orbits can be extended to general
\textit{stationary} axisymmetric asymptotically flat vacuum spacetimes. Hereafter, we assume
the radial pressure component of the surface stress-energy tensor $S_{ab}$ to
vanish (due to the choice $z_{\pm}=\pm b$). 
Models of `warm' disks with positive radial pressure is studied in \cite{Pichon96} where material of the 
disk is described by the distribution function of orbiting particles with various orbital parameters 
such that their averaged radial motion generates the radial pressure. Clearly, if such distribution function exists, the energy conditions for stress energy tensor
(see eqs. (3.12a-d) in \cite{Pichon96}) must hold. However, to get a distribution function
describing the population of particles with given orbital parameters requires the solution of
complicated integral equations. 

As stated above, we assume $p_r=0$. For positive angular pressure, $p_{\phi'}>0$, the distribution
of orbits is given by a simple model of two streams of counter-rotating circular geodesic dust particles.
The non-diagonal metric implies the non-diagonal surface stress-energy tensor and the counter-rotating 
streams can no longer have identical surface densities and 
equal opposite angular velocities. We assume the disk to be composed of two streams of dust particles 
moving in  the plane of identification $\Sigma$ with velocities $U_{\pm}^\mu
= N_\pm [1, \Omega_{\pm},0,0]$ and densities $\mu_\pm$, i.e., our stress-energy tensor has the form 
\begin{equation} \label{defS}
S^{\mu \nu} = \mu_+ U^\mu_+ U^\nu_+  +
                   \mu_- U^\mu_- U^\nu_- .
\end{equation}
Here $N_\pm^2=-1/\left( g_{{\phi \phi}}\Omega_\pm^2 + 2 g_{{t \phi}} \Omega_\pm + g_{{t t}}\right) $.
Both streams are made of particles satisfying geodesic equations
which due to symmetries simplify to
\begin{equation}
\label{Upm_geod}
 U^\mu_\pm U^\nu_\pm g_{\mu\nu,\rho}  =     0.
\end{equation}
Eqs. \eqref{defS} and \eqref{Upm_geod} represent 
five equations for four unknown quantities $\mu_\pm$ and $\Omega_\pm$.
In an analogy with the static case, the stress-energy tensor is restricted by the Einstein equations (Bianchi identities) 
so that $S^{\mu\nu}_{~~;\nu}=0$  (see e.g. \cite{Kuchar68}).
For stationary axisymmetric metric this simplifies to $S^{\mu\nu} g_{\mu\nu,\rho}=0$ and
the fact that this linear combination of stress-energy tensor components vanishes means that among three equations \eqref{defS} only two are independent. 

We thus first solve the radial component of the geodesic equation for the circular geodesic,
\begin{equation}\label{geoOmega}
g_{{\phi \phi,\rho}}\Omega_\pm^2 + 2 g_{{t \phi,\rho}} \Omega_\pm + g_{{t t,\rho}} = 0,
\end{equation}
to determine the stream velocities.
Both co-rotating and counter-rotating angular velocities $\Omega_\pm$
must be checked whether they imply time-like vectors.
If two subluminal circular geodesics exist, their
orbital frequencies $\Omega_\pm$ can be substituted into \eqref{defS} from which we find
\begin{equation} \label{mu_pm}
\mu_\pm = \pm\left( g_{{\phi \phi}}\Omega_\pm^2 + 2 g_{{t \phi}} \Omega_\pm + g_{{t t}}\right) {
{\Omega_\mp S^{tt} - S^{\phi t} }
\over
{\Omega_+ - \Omega_- }
              } .
\end{equation}

In CRGS model, only positive stream densities $\mu_\pm$ are acceptable.
The positivity of the stream densities $\mu_\pm$ is related to
the sign of the determinant of the 2x2 matrix $S^{AB}$
($A,B=t,\phi$), because
\begin{align}\label{detSAB}
 \det S^{AB} &= \rho^{-4} \det S_{AB} = N_+^2 N_-^2 \mu_+ \mu_- (\Omega_+-\Omega_-)^2
 \nonumber\\
 &= \mu p_{\phi'} (U^t V^\phi - U^\phi V^t)^2,
\end{align}
with $U^\mu = e^\mu_{(t')}, V^\mu = e^\mu_{(\phi')}$.
Using the standard criteria for positivity of quadratic form $S^{AB}$ we can state:
Both stream densities ${\mu}_\pm$ are {positive},
if $S^{AB} X_A X_B > 0$ for at least one 
2-component vector $X_A$, $\det S^{AB}>0$ and
two subluminal solutions $\Omega_\pm$ of (\ref{geoOmega}) exist.
Namely, if RR model yields disk with positive $\mu$ but negative pressure $p_{\phi'}$ somewhere,
even if both circular geodesic exist at these radii, one of the counter-rotating streams must have a negative density.

In \cite{BLPRL93} we introduced another model of counter-rotating streams -- working in the diagonal frame
\eqref{FIOtetrad}, we divided the ring density into two equal-density streams with opposite velocities
providing the right value of the pressure $p_{\phi'}$. We did not assume that particles creating
these two streams are in geodesic motion so the necessary condition for this interpretation was
$\mu > p_{\phi'} > 0$. As this model assumes some exchange of radial momentum between streams to keep
them in circular orbits, in the present work we will prefer the CRGS model of geodesic streams
of unequal densities.

The $z$-component of the geodesic equation determines whether a particle occurring slightly off the disk is 
returned back to the plane of the disk and is thus relevant for the stability of an individual circular
orbit. However, we do not address the problem of stability here.

\section{Tomimatsu-Sato spacetimes}
Tomimatsu-Sato (TS) solutions of the Einstein equations were constructed as the generalization of the Kerr solution. Soon after Kerr's discovery, Ernst showed that the Einstein equations for stationary axisymmetric field are equivalent to a single non-linear elliptic equation for a complex potential $\xi$ which for Kerr black-hole solution takes extraordinarily  simple form $\xi=1/(px+iqy)$ in spheroidal coordinates $x,y$. (These are simply related to Boyer-Lindquist coordinates, the parameter $q$ 
relates to black hole rotation and $p=(1-q^2)^{1/2}$. More detailed explanation follows.)
Tomimatsu and Sato then showed that there are other solutions of the Ernst equation in the form of complex rational function of $x,y$ representing field of a localized source \cite{TomimatsuSato}. They are labeled by integer parameter  $\delta = 1,2,3, ...$ ($\delta=1$ is Kerr) and the Ernst potential is the quotient of polynomials of degree $\delta^2-1$ and $\delta^2$.
This yields asymptotically flat spacetimes with mass $M$, 
angular momentum $J$ and the quadrupole mass moment $Q$ \cite{TomimatsuSato} given by
\begin{align} 
\label{TSJQ}
J=M^2\,q, ~~~ Q = M^3\left(q^2+\frac{\delta^2-1}{3\delta^2}(1-q^2)\right).
\end{align}
The metric potentials appearing in \eqref{WeyPap} are then determined by rational functions 
\begin{align}
e^{2\nu} = \frac{L}{E},~~~~
e^{2\zeta}=\frac{L}{F},~~~~ {\rm and}~~~~
A=\frac{B}{L},
\end{align}
where $L, E, F$ and $B$ are real-valued polynomials in $x, y$.

\begin{figure}[b!]
\centerline{\includegraphics[width=8.5cm]{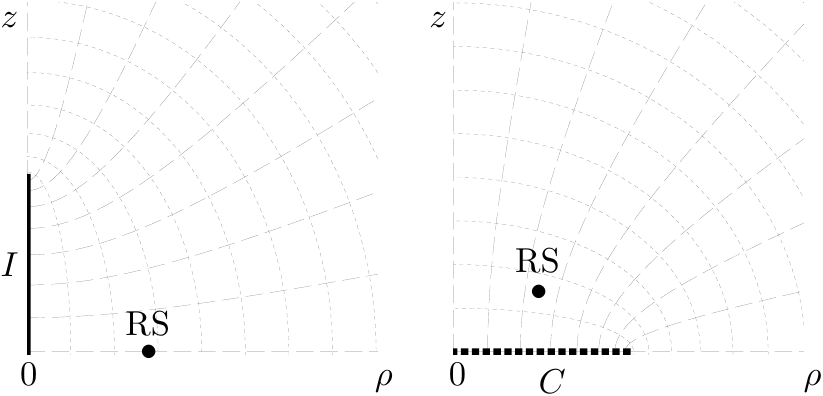}}
\caption{
\label{XYcoords}
The prolate (left) and oblate (right) spheroidal coordinates in the $\rho-z$ plane of the Weyl-Papapetrou coordinates. Together with lines of constant coordinates $x$ (dashed lines) and $y$ (long-dashed line), 
coordinate singularities $I$ and $C$ are plotted.
Depending on the value of TS family parameter $\delta$ the line $I$ represents a horizon or
a rotating conical singularity with horizons on each end. The cut $C$ can be opens into the coordinate patch $x<0$. Generic position of ring sigularities for $\delta=2$ is shown. See text for more details. 
}
\end{figure}

To express $e^{2\nu},e^{2\zeta}$ and $A$ in the form  of rational functions, spheroidal coordinates $x, y$ 
are needed. They are related to Weyl coordinates $\rho, z$ by transformation
\begin{align} 
z  = & {M p \over \delta} x y, 
~~~~~~~~~~
\label{DEFzrho2}
\rho = 
{M p \over \delta} \sqrt{1-y^2}\sqrt{x^2-\kappa }.
\end{align}
This prescription unifies prolate ($\kappa=+1, |q|<1$), oblate ($\kappa=-1, |q|>1$) spheroidal coordinates  and spherical coordinates ($\kappa=0, |q|=1$). The transformation introduces the scale proportional to the mass $M$ leaving the coordinates $x,y$ dimensionless. For a rotation parameter $q$ give, the value of the associated parameter $p$ is a positive solution of an equation $\kappa p^2+q^2=1$ (choosing $p=1$ for $q=\pm 1$). The introduction of the parameter $\kappa$ also unifies the expressions for metric potentials in a way which is equivalent to complex transformation $p\rightarrow -i p$ used in \cite{TomimatsuSato}. For $\delta=1$ there is a simple relation between the spheroidal coordinates $x, y$ and the Boyer-Lindquist coordinates $r, \vartheta$ \cite{BoyLindq67} 
\begin{align}
   r  =  M (p x + 1),~~~~
   \cos \vartheta =  y~.
\end{align}
The WP line element in these unified coordinates \eqref{DEFzrho2} reads
\begin{align} 
 d{s}^2 = &- e^{2\nu} (dt+A d\phi)^2 + e^{-2\nu}  \rho^2 d\phi^2
 \nonumber\\
 &+\frac{M^2 p^2}{\delta^2}\frac{x^2 - y^2 \kappa}{e^{2\nu-2\zeta} }\left(\frac{dx^2}{x^2-\kappa}+\frac{dy^2}{1-y^2}\right).
 \label{ds2xy}
\end{align}
The key properties of the TS metrics and coordinates \eqref{DEFzrho2} are schematically shown in Fig \ref{XYcoords}. 

In the stationary WP coordinates \eqref{WeyPap} the Tomimatsu-Sato spacetimes satisfy the vacuum Einstein equations everywhere apart from singular regions of coordinates \eqref{DEFzrho2} and certain singular rings. These ring singularities (RS), must be located at roots of the polynomial $E$ (see, e.g., the discussion based on the structure of the curvature invariants in \cite{Kodama}). 
As a consequence of the construction from a complex Ernst potential, 
the polynomial $E$ has the form of a sum of squares of two polynomials, $E=E_1^2+E^2_2$; so singularity $E=0$ occurs at the intersection of curves $E_1=0$ and $E_2=0$. Thus, in $x-y$ or $\rho-z$ plane these singularities are points rather than lines. Then, considering also the angular coordinate $\phi$ we get the ring singularities.

For a {\bf small rotation parameter} $|q|<1$ the TS metrics require {\it prolate} spheroidal coordinates. These become singular at $x=1$. The surfaces of constant $x$ are confocal ellipses in $\rho-z$ plane which
for $x=1$ reduce into a segment $I$ of the $z$-axis, $z\in [-Mp/\delta,Mp/\delta]$, and 
the spacetime character of both Killing vectors at the limiting hypersurfce $I$ changes with $\delta$.
For the Kerr black hole ($\delta=1, |q|<1$) the segment $I$ is the event horizon $r=r_+=(1+p)M$ and the prolate spheroidal coordinates cover the Kerr spacetime `above' the outer horizon. Similarly, for $\delta=3$ the segment $I$ is a rotating horizon with regular values of Weyl invariants \cite{Hoenselaers1979}.  The Killing vector $\partial/\partial \phi$ is timelike on $I$ for  $\delta$ even. Various analysis of structure of $I$ for $\delta=2, |q|<1$ took many years\cite{TS72PRL,Gibbons}, until \cite{Kodama} showed that the interior of $I$ behaves as a `rotating' conical singularity (known from NUT spacetime) and endpoints of $I$ turn out to be shrunken horizons of spherical shape.  

Apart from $I$, for $\delta=2,3$ there is a ring singularity located at the equatorial plane of WP coordinates. Other ring singularities are present in a detached $x<-1$ patch of prolate spheroidal coordinates which we do not consider here (in Kerr black hole spacetime, this is the well-known RS in the region $r<r_-$).

Another feature appearing in stationary axisymmetric spacetimes are regions of CTCs at places where the
Killing vector of the axial symmetry becomes timelike, $g_{\phi\phi}<0$. In the TS class, only Kerr black hole spacetimes are free of CTSs in the domain of outer communication. Even with $|q|<1$ for $\delta>1$ there are toroidal CTC regions (see Fig. \ref{FIG_TS32_08}).

For $|q|=1$ any $p>0$ can be chosen and all TS metrics coincide with the extreme Kerr metric.
For a {\bf large rotation parameter} $|q|>1$, $\kappa=-1$, {\it oblate} spheroidal coordinates are required, for which the transformation \eqref{DEFzrho2} 
of the domain $x\in I\!\!R, -1 \le y\le 1$ covers twice the $\rho-z$ half-plane, 
where the sheet $x<0$ has an asymptotically flat infinity with negative mass.
For the Kerr spacetime the only RS is located in the equatorial plane $y=0$ of $x<0$ sheet. For TS with $\delta$ larger there are also symmetrically placed off-equatorial pairs of ring singularities. 
With $|q|>1$ there are two RS with $y=0, x<0$ and one pair of RS with $x>0$ for $\delta=2$ and 
three RS with $y=0, x<0$, one pair of RS with $x<0$ and 2 pairs of RS with $x>0$ for $\delta=3$.
The location of these RS is not studied in literature, because only
$|q|<1$ sources are discussed in detail, even though, unlike in the Kerr case $\delta=1$, no cosmic censorship argument favors the $|q|<1$ case for $\delta>1$.)

While in the construction of disks without the radial pressure we stay in the $x>0$ region of coordinates, one can in priciple enterthe $x<0$ region if disk with non-zero radial pressure are admitted. Unlike in the $|q|<1$ case where regions of positive and negative $x$ are in WP coordinates disjoint, for $|q|>1$ the WP metric is regular on the cut $x=0$. This has the form of disk $C:=\{[\rho,\phi,z];0\le\rho<Mp/\delta,\phi\in[0,2\pi), z=0\}$ in WP coordinates. 
This cut can be understood either as a finite-radius disk source of the given TS spacetime
(however, in Section VI we show that it would then be made of unphysical matter), or as an entrance into the region of negative $x$. Some features appearing in the $x>0$ as an ergosphere (which can be produced by physical disks) appear to originate at singularities present in the negative $x$ sheet. This we illustrate for the TS solutions with $\delta=1,2$ in Fig. \ref{Znegative} where a transformation smoothly covering $C$ is used.  Among others, location of regions of CTCs in the $x<0$ sheet of Kerr spacetime is shown.
In this figure as well as in Figs. \ref{FIG_KERR}, \ref{FIG_TS2}, \ref{FIG_TS32_08} and \ref{FIG_TS32_12} we use the Weyl-Papapetrou coordinates $\rho$ and $z$ scaled in units of $M$.

\section{Properties of disk sources}
\begin{figure}[!htb]
\centerline{
\includegraphics[width=8cm]{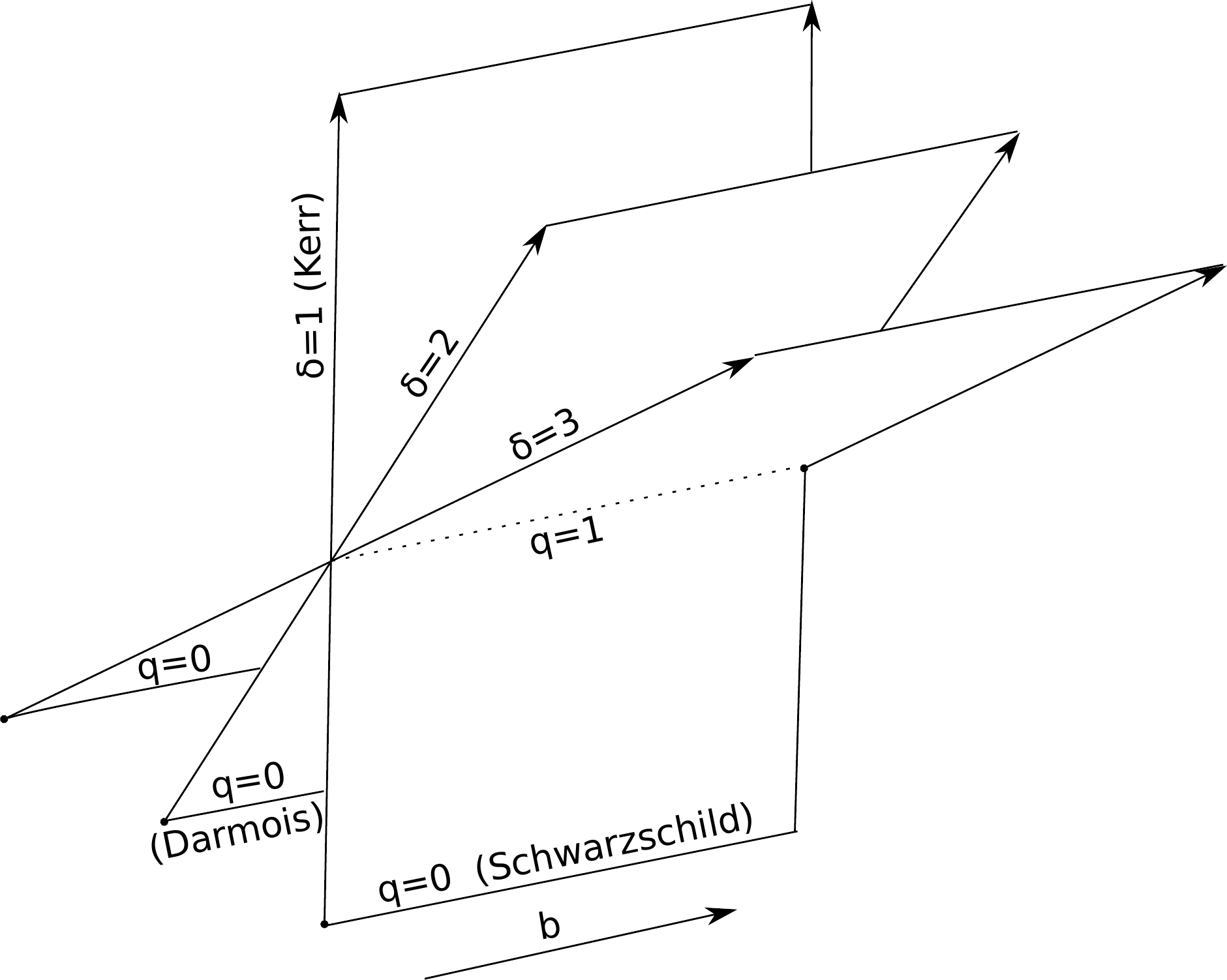}
} 
\caption{
The schematic picture of the parameter space of disk spacetimes considered.
Each disk is characterized by an integer $\delta=1,2,3$ and
by relative angular momentum $q=J/M^2$ (both are the parameters of the original spacetime) and by the
size $b$ of the cut-off region. The parameter $b$  
determines the compactness of the disk. The mass parameter $M>0$ is a simple parameter of scale,
and one can put $M=1$. The extreme limit $q=1$ is the same for any $\delta$. 
\label{FIG2}}
\end{figure}

In the following we study the disks (without radial pressure so $\Sigma^+$ is the hypersurface $z=b$) 
as sources of the Tomimatsu-Sato spacetimes with mass $M$, rotation parameter $q$ and an integer distortion parameter $\delta=1,2,3$; the parameter space is illustrated in Fig.\ref{FIG2}.

Within this parameter space we explore whether:
\begin{enumerate}[(i)]
 \item the horizon, singularities or closed timelike lines (CTCs) are contained inside the removed region;
 \item energy conditions \eqref{wecdiag} or \eqref{decdiag} hold;
 \item the pressure $p_{\phi'}>0$ for RR model;
 \item real solutions for the circular geodesics of \eqref{geoOmega} exist;
 \item velocities given by \eqref{geoOmega} are subluminal.
\end{enumerate}
First, the following implications are useful to be noticed:
(iii)$\implies$(ii) and (v)$\implies$(iv). From \eqref{ECimplicatons} we can also deduce that 
\begin{enumerate}[(vi)]
 \item in RR model: (ii)$\implies \mu>0$;
 \item in CRGS model: (ii)$\wedge$(iii)$\wedge$(v) $\implies \mu_\pm>0$.
\end{enumerate}

We also display the properties of the disks described by specific parameters by constructing the
plots of the radial dependence of various physical quantities. As a radial coordinate in such plots we use the circumferential radius $R={g_{\phi\phi}^{1/2}}$. 

To illustrate various densities in the situation when disks and spacetime geometries are very different 
from a flat geometry,
we adopt two approaches. First, we plot cumulative distribution profiles which show what fraction 
of the Komar mass or of the angular momentum is generated by the central part of the disk 
below some radius $R$.
Then, we plot the radial profiles of the densities appearing within the Komar integrals.
For our disks with $z(s) = b$, we write integrals \eqref{KomarMDX} and \eqref{KomarJDX} over the surface densities $\mu_D, j_{D}$ 
defined with respect to the circumferential radius $R$ as:
\begin{align}\label{def_muD}
M_D &= \int_0^\infty\!\!\! \mu_D~2\pi{ R}~d{ R}, ~
\mu_D= {2 \rho\,{g_{\rho\rho}^{1/2}}\over g_{\phi\phi,\rho}}
\left(S^\phi_\phi-S_t^t\right),
\\\label{def_jD}
J_D &= \int_0^\infty\!\!\! j_D~2\pi{ R}~d{ R}, ~
j_D= {2 \rho\, {g_{\rho\rho}^{1/2}}\over g_{\phi\phi,\rho}}S^t_\phi.
\end{align}
This approach enables us to assign also the densities $\mu^\pm_D$ to both counter-rotating streams 
in \eqref{defS}; their radial integrals then determine how large fraction of the total (Komar) disk mass
is generated within each stream.

\subsection*{The disk sources of the Tomimatsu-Sato spacetimes}
In this section we use the general results given above to construct the disk sources of the Tomimatsu-Sato spacetimes.
Let us define the ``derivative operation'' denoted by ``$\prime$" acting on a function $f(x,y)$ as
\begin{align}
f' & =  (x^2-\kappa ) \frac{\partial f}{\partial x} +  (1-y^2)\frac{x}{y} \frac{\partial f}{\partial y}.
\end{align}
The $\partial_z$ derivative appearing in the general formula for the surface stress tensor 
\eqref{Sabzkonst} can be expressed as  
\begin{align}
\frac{\partial f}{\partial z} & =  \frac{\delta}{M p} \frac{ y}{x^2-\kappa  y^2}  f'.
\end{align}
Then, employing this operation, we find the  surface stress-energy tensor components \eqref{Sabzkonst}
in the TS metrics to 
read
\begin{align} 
 S_{tt} &= {ZLF \over E^2} \; \left({L^\prime \over L} + {F^\prime \over F} - 2
\, {E^\prime\over E}\right), \\
 S_{t\phi} &= {ZBF \over E^2} \; \left({B^\prime \over B} + {F^\prime \over F} - 2\,
{E^\prime \over E}\right),\\
 S_{\phi\phi} &= {ZF \over L} \; \Big[ {\left(\rho^2 - \frac{B^2}{E^2}\right)} \left({L^\prime
\over L}-{F^\prime \over F}\right) 
\nonumber\\
&~~~~~~~~~~~~~~+ 2\, {B^2 \over E^2}\left({B^\prime \over B}-
{E^\prime \over E}\right)\Big], 
\end{align} 
where 
a substitution has to be made for $x$ and $y$ from \eqref{DEFzrho2} so that the expressions become functions of the coordinate $\rho$ in the plane $z=b$ of the disk, and
\begin{align}
Z = {1 \over 8\pi} \frac{\delta}{M p} \; \sqrt{{E \over F}} {y\ \over x^2 - \kappa y^2}.
\end{align}

For vacuum spacetimes the only realistic case in the TS family is the Kerr black hole spacetime with $|q| \le 1$, otherwise naked singularities occur in the central regions. 
In our approach central parts are replaced by a disk and the value of $q$
itself is not directly restricted by the presence of naked singularities in the complete solution.
Instead, the size of the region between $\Sigma^-$ and $\Sigma^+$
that must be excluded is determined by the properties of the disk, namely, whether the energy conditions
hold or if the cut-off region admits the particular model of the disk matter (such as RR or CRGS).

For the Tomimatsu-Sato spacetimes with small angular momentum $J$ and identifications performed at large $b$, both the disk sources and the fields will be weak and similar to the Schwarzschild-disk cases \cite{BicakLyndKatz93}. For such disks both RR and CRGS interpretations are possible.

The CRGS model requires the positive value of the determinant \eqref{detSAB} and this condition may be broken even in the weak field limit $b\gg M$, when pressures are still small, but negative.
Using the expansion of the TS metric functions we can find an approximate expression for \eqref{detSAB} and
check its sign. It turns out that the spacetime geometry prohibits CRGS disk sources for large enough
angular momentum since negative pressures $p_{\phi'}$ appear near the axis.
In this approximation both CRGS densities are positive only if  
$b/M\gtrapprox 2+9q^2/4 \gg 1$, or, expressed in terms of the disk mass and angular momentum, 
\cite{TomimatsuSato}
\begin{align}
b\gtrapprox 2M+\frac{9 J^2}{4 M^3} \wedge b\gg M. 
\label{bMinApprox}
\end{align}
In the higher order approximation the quadrupole moment (which depends also on $\delta$) enters this formula. 

\begin{figure}[b!]
\centerline{\includegraphics[width=7.5cm]{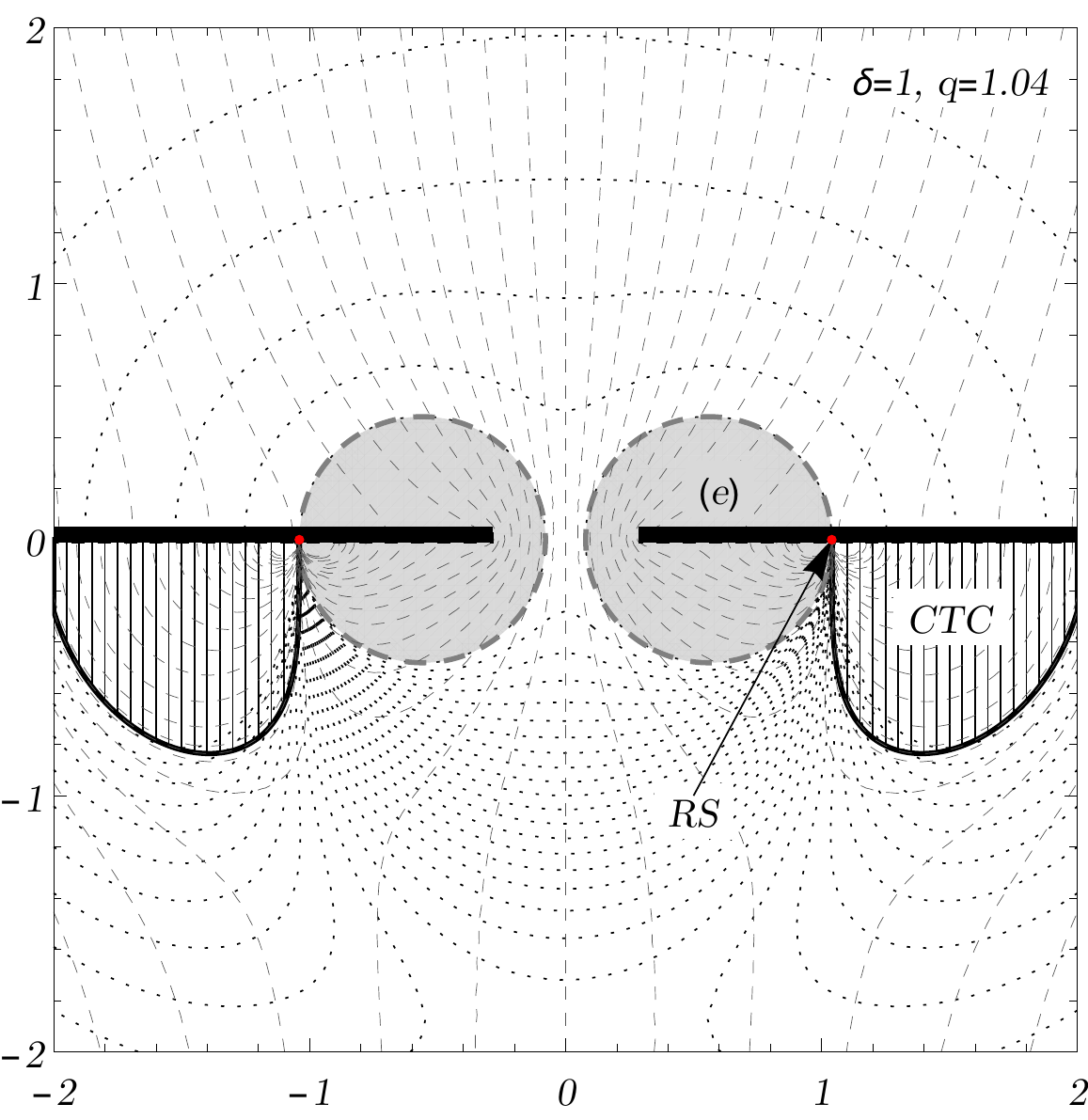}}
\centerline{\includegraphics[width=7.5cm]{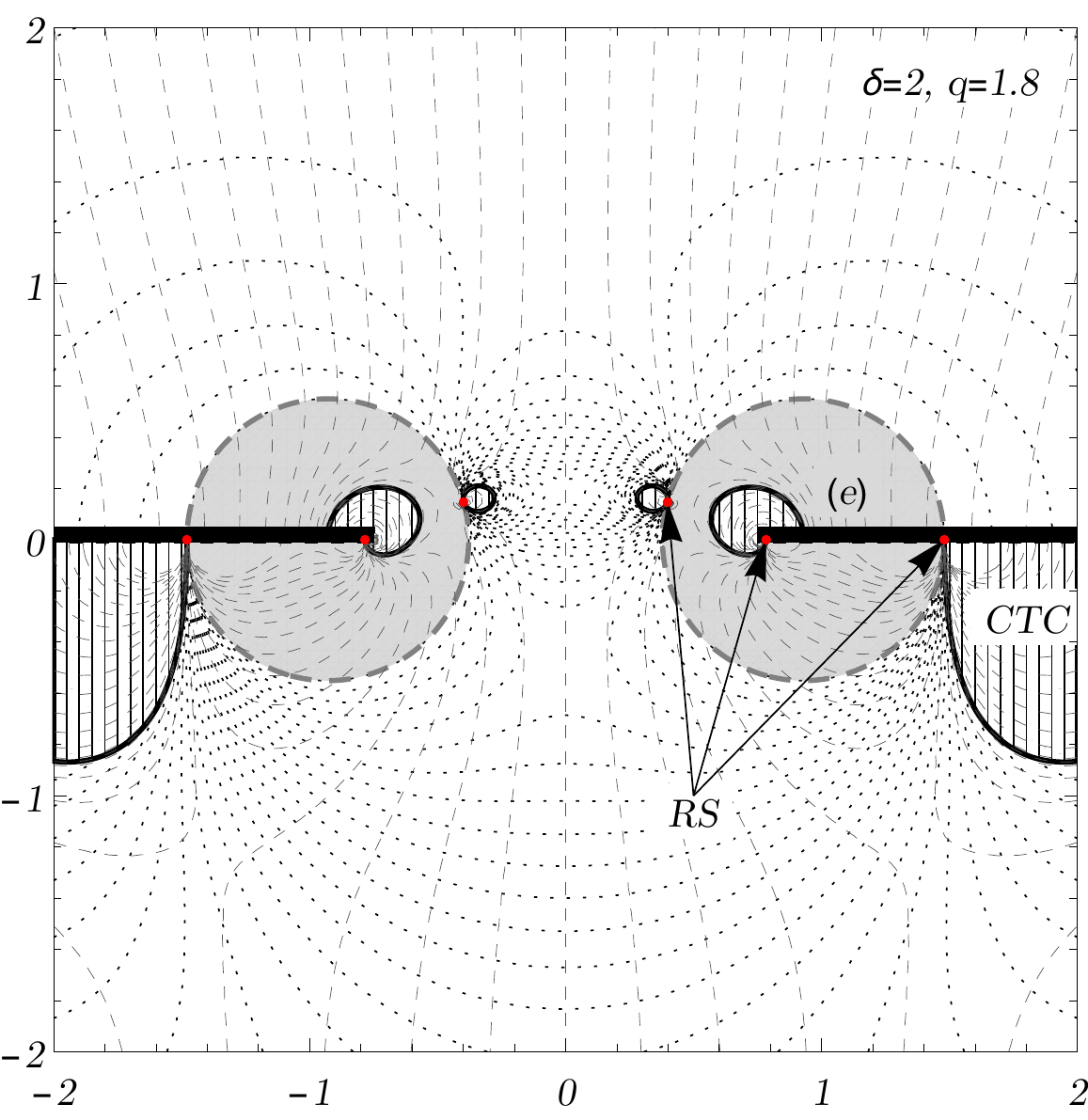}} 
\caption{
\label{Znegative}
For $|q|>1$ the oblate spheroidal coordinates can be simply extended to the patch $x<0$. When plotted in $\rho-z$ plane in the Weyl-Papapetrou coordinates
this extension yields a cut in $z=0$ plane along $\rho>M p/\delta$ (thick horizontal line) so that
the standard region ($x>0$) is in the upper part and 
extended ($x<0$) region is in the lower part of the plot. 
To show rotational symmetry, we extend $\rho$ symmetrically in the left half of the plot.
The toroidal ergosphere (e), CTC regions and ring singularities of $|q|>1$ Kerr  (top) and TS $\delta=2$ (bottom) spacetimes are illustrated in this figure.
The identifications along $z=\pm b$ can provide disk sources with ergoregions for the Kerr case, and for $\delta>1$ disks can also be surrounded by toroidal regions of CTC. Contours show norms of Killing vectors, $g_{tt}=-0.1,-0.2,...$ (dotted) and $R=g_{\phi\phi}^{1/2}=0, 0.25, 0.5,...$ (dashed).
}
\end{figure}

Since the Kerr solution is the member of the TS family we should mention some implications which the last 
conditions bear on the cosmic censorship conjecture. This admits only the formation of the Kerr black holes
which have $|q|\le 1$. Although the construction of the disk sources does not in principle prevent us from
constructing the TS spacetimes with $|q|>1$, there is a ''barrier`` which disfavors the construction of
even mildly compact CRGS disk sources with large $q$. If the ratio  $M/b$ is taken as a field-strength
indicator, in the Kerr case ($\delta=1$) we have $M/b\lessapprox 4/(9 q^2)$. In this sense, a 
``parametric collapse'' of a CRGS disk into a naked Kerr singularity is prohibited already at early stages.

As $b$ becomes smaller, the relativistic effects play more important role and 
at some point truly strong-field features may prevent the disk source to have physically plausible
properties: one of circular Keplerian geodesics may become superluminal, 
or the violation of energy conditions occurs (either $S^{ab}$ cannot be diagonalized or energy density
of the RR becomes negative at some radii). For the set of spacetimes we consider,
it was confirmed that unless the energy conditions are violated, 
{\it closed timelike curves do not exist in the spacetime surrounding the disk}. 
Disks compact enough, with parameters close to extreme black holes, exhibited the presence
of the ergoregions in the vicinity of the disk but because at least one of 
$\Omega_\pm$ is tachyonic there, {\it only disks made of RR
can be surrounded by a toroidal ergoregion}.

As shown in Fig. \ref{Znegative} for
Tomimatsu-Sato $q>1$ spacetime, the identification along the surface $z=\rm const.$ could in principle reveal parts of the spacetime 
with closed time-like curves (CTC) extending to a distant observer. 
The extension into negative values of the coordinate $z$ shown in this figure for $|q|>1$ are discussed in  e.g. \cite{Carter1968}, or more recently, in \cite{ZadehJMP2015} and \cite{Brauer2015}.
In the following sections we show that for TS $\delta=2,3$ spacetimes the energy conditions 
are violated already before the CTC domain appears around the disk.
It illustrates certain `parametric form' of the chronology protection in general relativity.

\begin{figure}[!htb]
\centerline{\includegraphics[width=7.5cm]{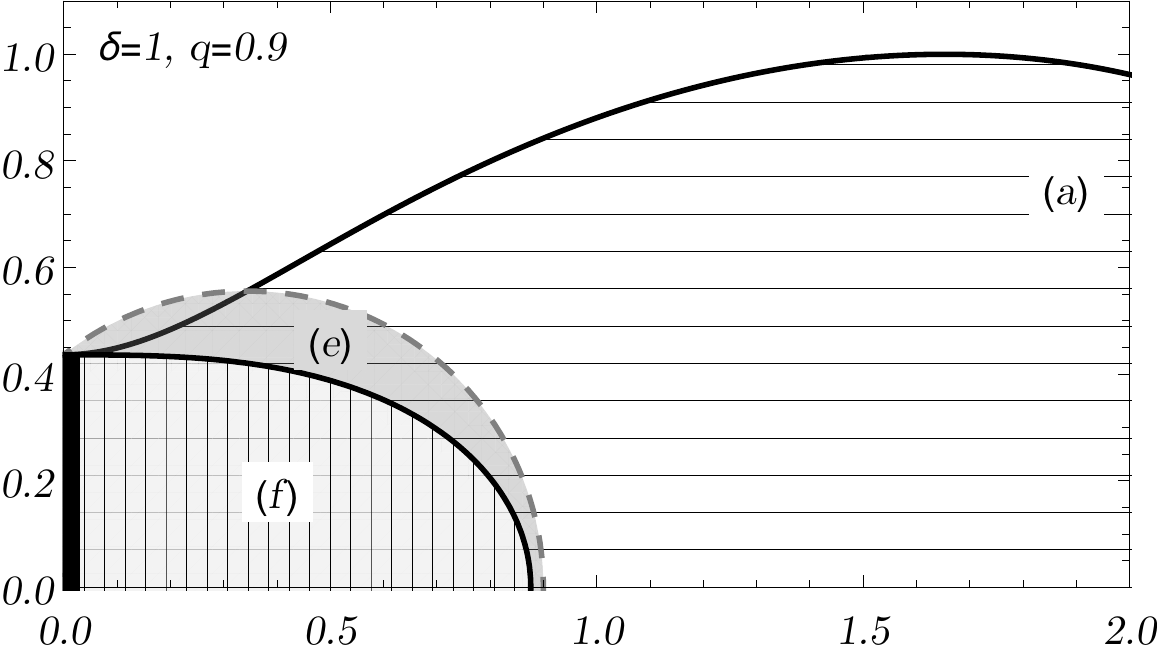}}
\centerline{\includegraphics[width=7.5cm]{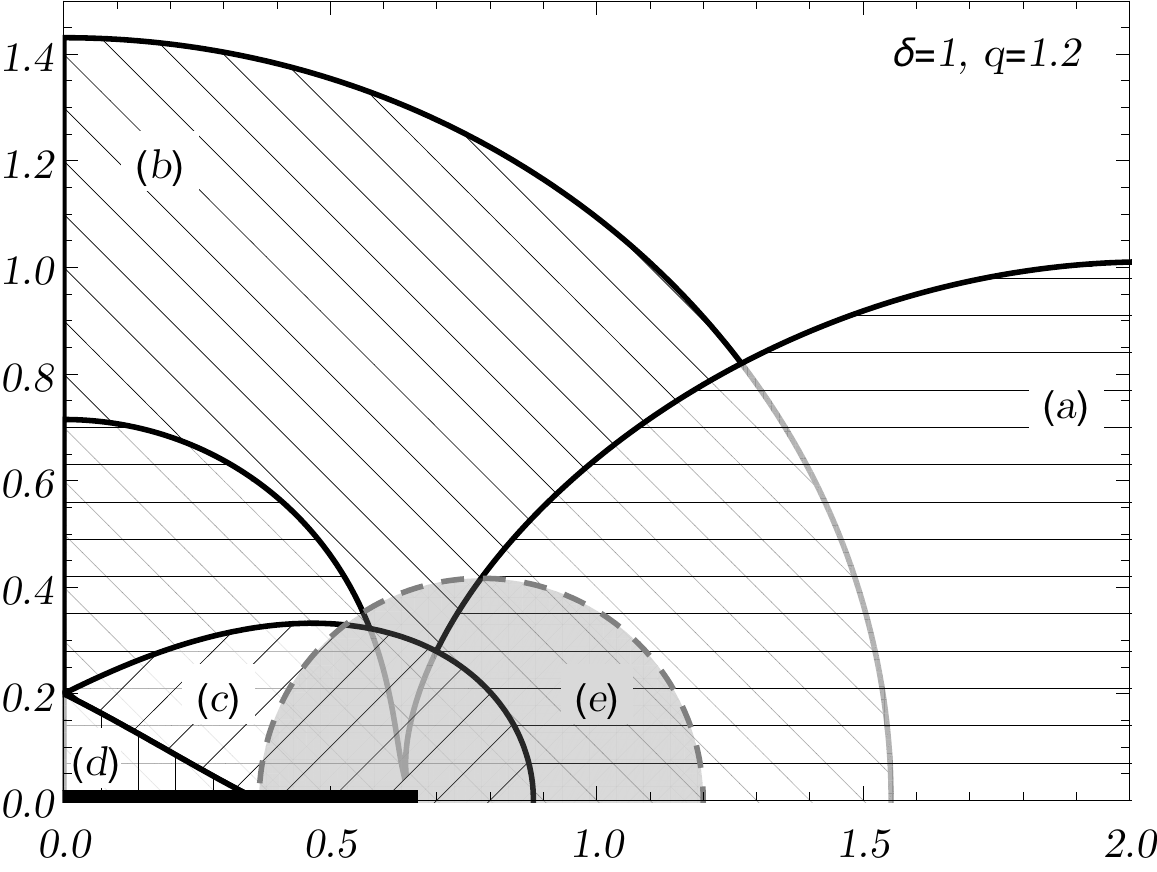}
} 
\caption{
Properties of the stress-energy tensor of a $p_r=0$ disk as a source of the Kerr geometry in the $\rho-z$ plane of Weyl-Papapetrou coordinates for $M=1$ Kerr black hole (top) and Kerr naked singularity (bottom). 
To find out these properties one draws a horizontal line $z=b$ (corresponding to the identification surface $\Sigma^+$) and along this line reads out the properties of the stress-energy tensor at given radius of the disk:
(a) One or both Keplerian velocities are superluminal;
(b) Negative value of ${\rm det}\;S_{AB} \sim \mu p_{\phi'} \sim \mu_+ \mu_-$;
(c) No diagonalisation is possible, $\sigma<0$;
(d) Energy density $\mu$ of RR is negative;
(e) Ergosphere;
(f) Pressure exceeding DEC limit, $p_{\phi'}>\mu$.
\label{FIG_KERR}
}
\end{figure}

\subsection*{Kerr ($\delta=1$) disks}
The most important solution in the TS family is the Kerr metric with the metric functions 
\begin{align}
 e^{2 \nu} &= \frac{p^2 \left(x^2-\kappa \right)-q^2 \left(1-y^2\right)}{(p x+1)^2+q^2 y^2},\\
 e^{2 \zeta} &= \frac{p^2 \left(x^2-\kappa \right)-q^2 \left(1-y^2\right)}{p^2 \left(x^2-\kappa  y^2\right)},\\
 A &= \frac{2 M q \left(1-y^2\right) (1+p x)}{p^2 \left(x^2-\kappa \right)-q^2 \left(1-y^2\right)}.
\end{align}

When the radial pressure-free disks with $\Sigma^+=\{x^\mu|z=b\}$ 
with $|q|\le 1$ are constructed as sources of the Kerr metric,
the excluded (``cut-off'') region must satisfy the condition $b>M p$ so that the horizon is completely
removed from the disk spacetime.
For $|q|>1$ the ring singularity lies in the $x<0$ patch of coordinates so any $b>0$ is in principle permissible. 
The singular region of the prolate spheroidal coordinates which coincides with the Kerr black hole horizon, and that of 
oblate spheroidal coordinates, through which patch $x<0$ could be reached, are shown as thick black solid lines in Fig. \ref{FIG_KERR}. This is plotted for the ``black-hole'' and ``naked Kerr''-rotation parameters $q=0.9$ and $q=1.2$, respectively and the properties of the disk stress energy tensor for various values of parameter $b$ and radial coordinate $\rho$ are shown.

{\bf Kerr black-hole metrics}. The properties of disk sources with $|q|<1$ are illustrated in the top panel of Fig. \ref{FIG_KERR}. One can construct RR disks with $|q|<1$ as long the whole black-hole horizon (thick vertical line in the figure) is contained between $\Sigma^+$ and $\Sigma^-$, i.e. $b>Mp$ must be chosen. If the disk would `cut into' the horizon, the radially ``static'' matter of RR would need pressure $p>\mu$ near the horizon.

For $1/\sqrt{2}<q<1$, spacetimes of compact enough RR disks may contain an ergoregion (light gray regions in the figure). The dashed region (a) shows places where two timelike counter-rotating circular geodesics in the plane $\Sigma^{\pm}$ do not exist and the disks cannot be made of CRGS. For $q<1$ this effect restricts the disk compactness to $b \gtrsim M$. 

{\bf Naked Kerr metrics}. For $|q|>1$ more obstacles to construct disk sources arise as can be seen in the bottom panel of Fig. \ref{FIG_KERR}. We already mentioned region (a) in which we do not have 
counter-rotating circular geodesics needed to construct CRGS surface stress energy tensor \eqref{defS}. For $a>M$ this region spreads also along the $z-$axis. Even if both geodesics exist, realistic CRGS models cannot have negative stream densities, which happens in region (b). For $q\gtrsim 1.12$ the positivity of stream densities restricts possible central redshifts more strictly than the existence of timelike CRGS geodesics. Still, the RR disk model is possible until the value of $b$ reaches the region (c). Then we cannot diagonalize the disk stress energy tensor and both WEC and DEC are violated.

\begin{figure}[!b]
\centerline{\includegraphics[width=7.2cm]{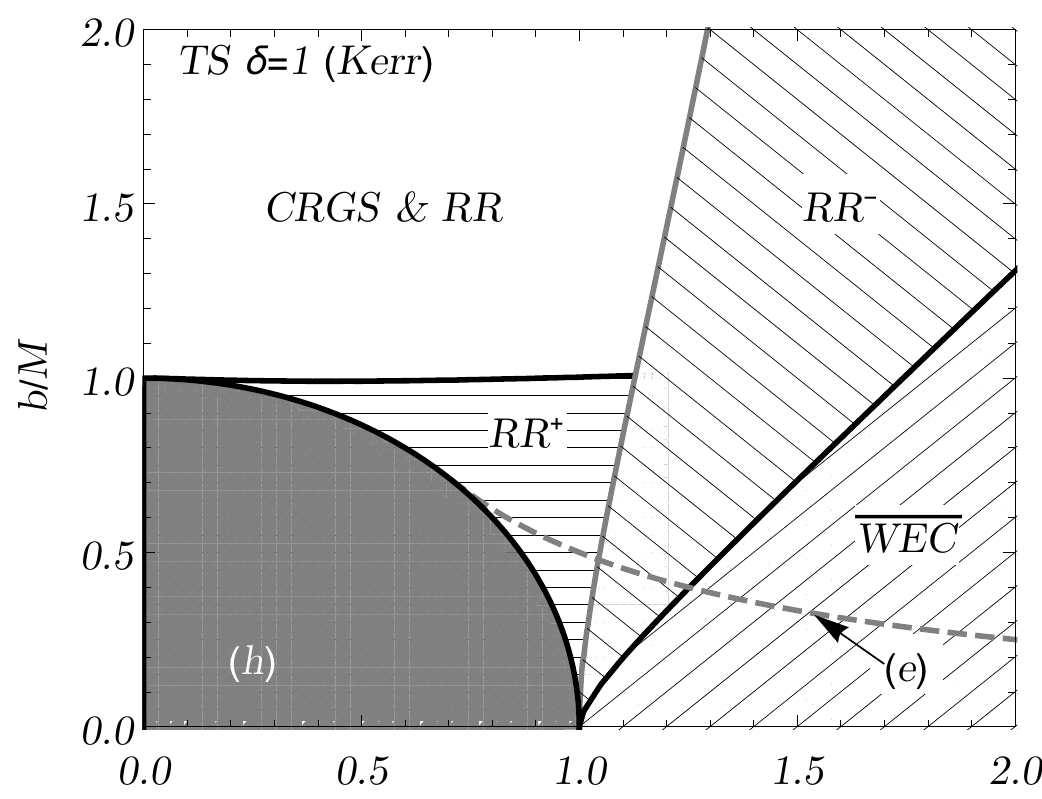}}
\centerline{\includegraphics[width=7.2cm]{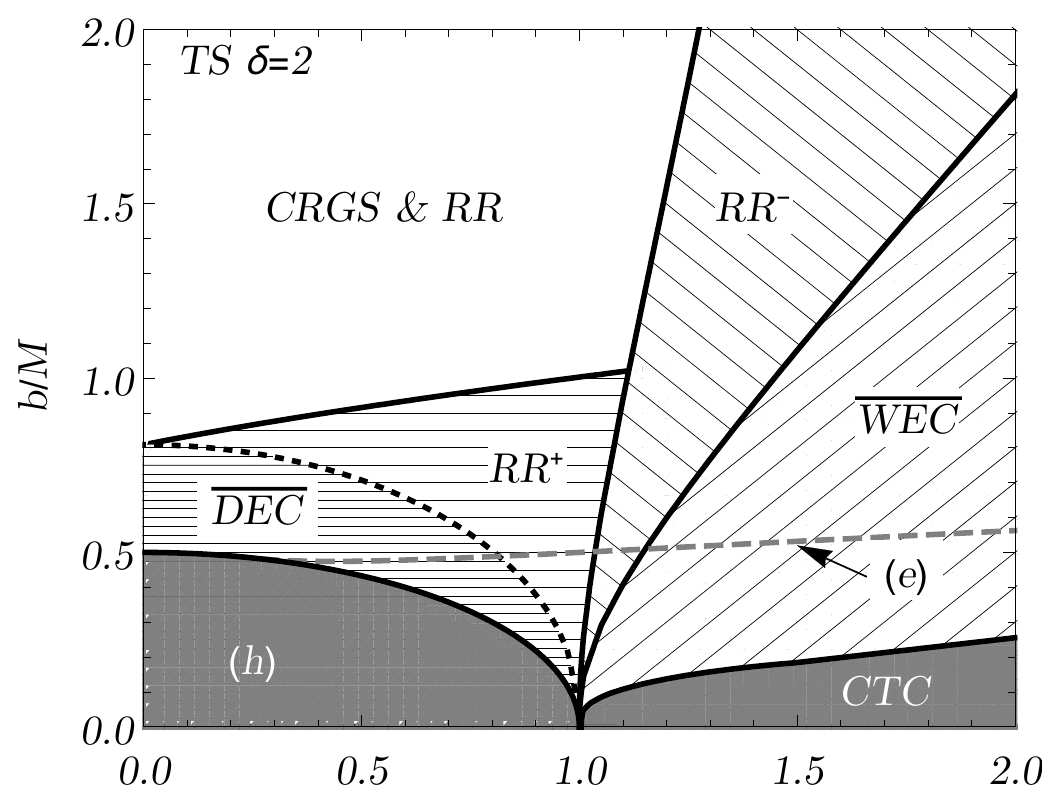}}
\centerline{\includegraphics[width=7.2cm]{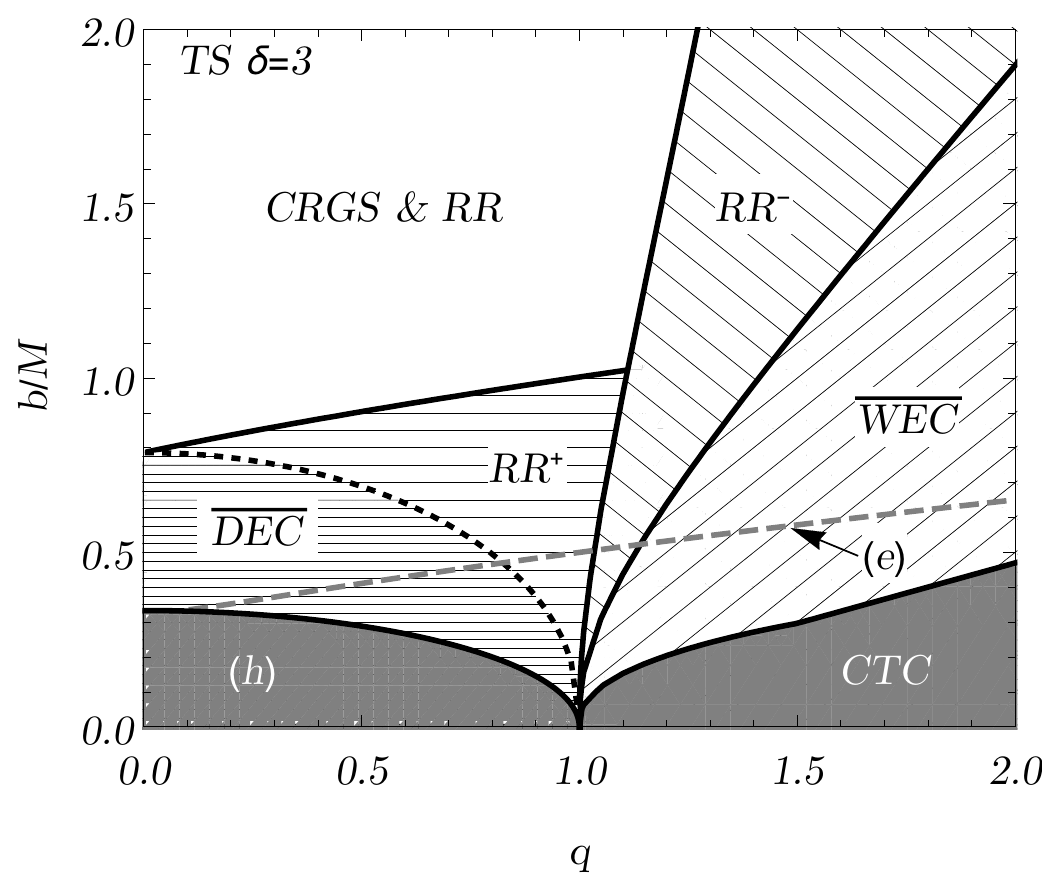}}
\caption{
\label{FIG_REGIO_KERR}
Properties of disks with given angular momentum parameter $q$ and
half-with of the excluded region $b$ generating TS metrics with $\delta=1,2,3$. 
For large enough $b$ both RR and CRGS models are applicable.
For given $b$ when increasing the angular momentum we enter $RR^-$ region where at some disk radii, negative tangential pressure appears. For smaller $b$ counter-rotating circular timelike geodesics no longer exists and we enter $RR^+$ region where RR have positive pressure everywhere. In $\overline{WEC}$ region of parameters, either the rings should have negative mass density or the stress-energy tensor cannot be diagonalized. In $\overline{DEC}$, pressure exceeds RR density at some radii. Disks with parameters below (e) have ergoregions, those in (h) or (ctc) would touch horizon or involve CTCs.
}
\end{figure}

For disks with even stronger fields (when $b<a-M$), we find that near the disk center a negative energy density arises, region (d).  
At the same threshold, $b<a-M$, a repulsive force starts to act on the particle at the disk centre. For the discussion of the repulsive effects of the naked Kerr singularity, see, e.g., \cite{Semerak1993}.
 
{\bf Disk compactness restrictions} due to the centrifugal effects are shown in
the top panel of Fig. \ref{FIG_REGIO_KERR}. 
There, for $\delta=1,2,3$, and for given $q$, we plot the minimal values of parameter $b$, for which the considered properties of the Kerr disk stress-energy tensor hold for $\rho\in [0,\infty)$. We analyze disks with $|q|<2$ and $b/M<2$ dividing the parameter space into the following regions indicated in 
Fig. \ref{FIG_REGIO_KERR}:
\\
\noindent  $\mathbf{RR\& CRGS}$: When large enough central region is removed by the identification, the stress-energy tensor can be interpreted using both RR and CRGS models. Their properties are discussed in detail in the following text. 
\\
\noindent  $\mathbf{RR^-}$: 
For given disk parameters, the RR model is applicable, but at certain disk radii there is a negative value of $p_{\phi'}$. Then, even if both circular geodesics exist, one of the stream densities would have to be negative as follows from Eq. \eqref{detSAB}. The upper boundary of the region is approximately described by the relation \eqref{bMinApprox}; an exact, but rather lengthy formula is given in \cite{BLPRL93}.
\\
\noindent  $\mathbf{RR^+}$: 
The RR model is applicable, both $\mu$ and $p_{\phi'}$ are non-negative everywhere. Disks cannot be made of CRGS because two timelike circular geodesics are not available at some radii.
\\
\noindent  $\mathbf{\overline{WEC}}$: 
At some disk radii the surface stress-energy tensor cannot be diagonalized, or the obtained energy density $\mu$ is negative. The weak and dominant energy conditions are violated.
\\
\noindent  $\mathbf{\overline{DEC}}$: 
At some radii $p_{\phi'}$ exceeds the energy density $\mu$, the dominant energy condition is violated. For $\delta=1$ this region is completely covered by (h).
\\
\noindent  $\mathbf{CTC}$:
The region excluded by the identification is not large enough to contain all CTC regions. Ring singularities may appear above the disk for even smaller $b$. (This can happen only for $\delta \ge 2$.)
\\
\noindent  $\mathbf{(h)}$: 
The region excluded by identification is not large enough to contain the horizon.
\\
\noindent  $\mathbf{(e)}$: An ergosphere occurs above the disk; this is possible for the RR disks with $q\approx 1$.

\begin{figure}[b!]
\centerline{\includegraphics[width=\linewidth]{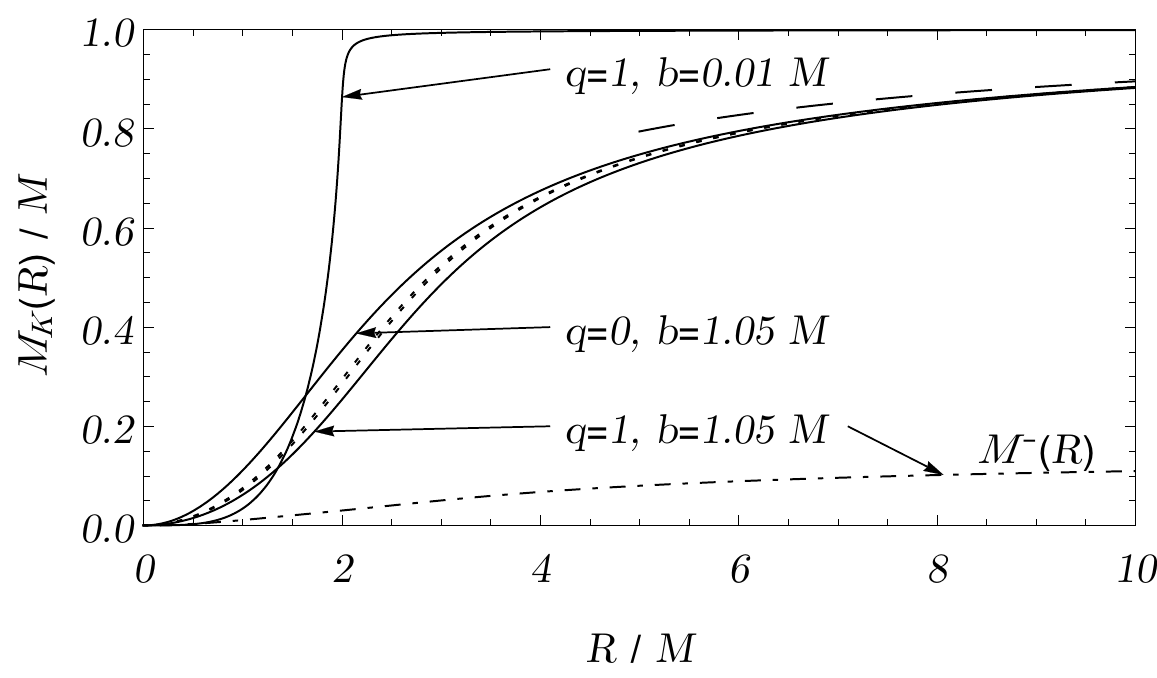}}
\caption{
Cumulative radial distribution of the Komar mass inside the Kerr disks as function of 
circumferential radius $R$. For $q\approx 1$, disks with a compact mass distribution are possible, for the CRGS 
models we have to choose $b \gtrapprox M$ and disk matter is spread farther out.
Mass fractions of the $b=1.05M$ Newtonian Kuzmin disk models are shown as a long-dashed line.
For $q=1,b=1.05M$, both CRGS have positive densities but $\mu^-$ component (dash-dotted line $M^-(R)$)
is responsible for only about 15\% of the total Komar mass of the disk. Dotted lines between $q=0$ and $q=1$ lines show profiles of $q=0, b=1.05M$ disks with $\delta=2$ and $\delta=3$ (top to bottom). 
\label{FIG_KerrCummu}
}
\end{figure}

Among the disks considered, only the Kerr $\delta=1$ disks for $|q| \le 1$  can have 
arbitrarily large central redshifts and satisfy energy conditions; these are the disks with
parameters $q,b$ at the boundary of region (h) in Fig. \ref{FIG_REGIO_KERR}.
The CRGS model yields arbitrarily large central redshifts only for the Schwarzschild
disk with $q=0, b \searrow M$.

\begin{figure}[b]
\centerline{\includegraphics[width=\linewidth]{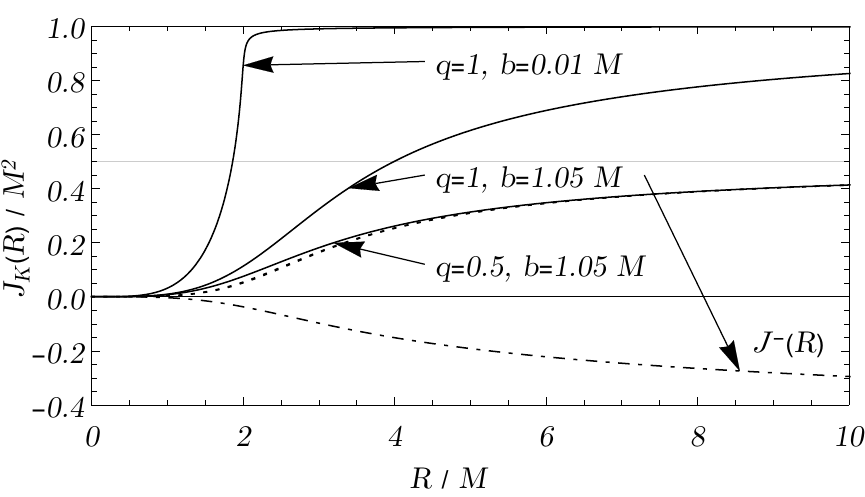}}
\caption{
Cumulative radial distribution of the Komar angular momentum inside the Kerr disks as a function of a 
circumferential radius $R$. For the parameters $q=1,b=1.05M$ a negative angular momentum due to $\mu_-$ carrying about $-57\%$ of the total angular momentum is also shown (dash-dotted line), remaining $157\%$
is deposited in the co-rotating stream. Dotted lines below $q=0.5$ line show profiles of $q=0.5, b=1.05M$ disks with $\delta=2$ and $\delta=3$ (top to bottom). 
\label{FIG_KerrAngularCummu}
}
\end{figure}

In Fig. \ref{FIG_KerrCummu} we illustrate the mass distribution inside the disks by expressing the relative 
mass contribution, $M_K(R)/M$, of the central part of the disk up to a given radius $R$, to the total Komar mass 
(19). Since the leading term of the asymptotic behavior  depends only on the mass parameter, 
at large radii we get the same profile as for the classical Kuzmin disks with $M_K(R)/M \doteq 1-b/R$, independently of 
the parameters $\delta$ and $q$. The distribution changes only slightly with the TS parameter $\delta$ 
as is also illustrated in Fig. \ref{FIG_KerrCummu}.
The distribution of the angular momentum is shown in Fig. \ref{FIG_KerrAngularCummu}.

\begin{figure}[b]
\centerline{\includegraphics[width=\linewidth]{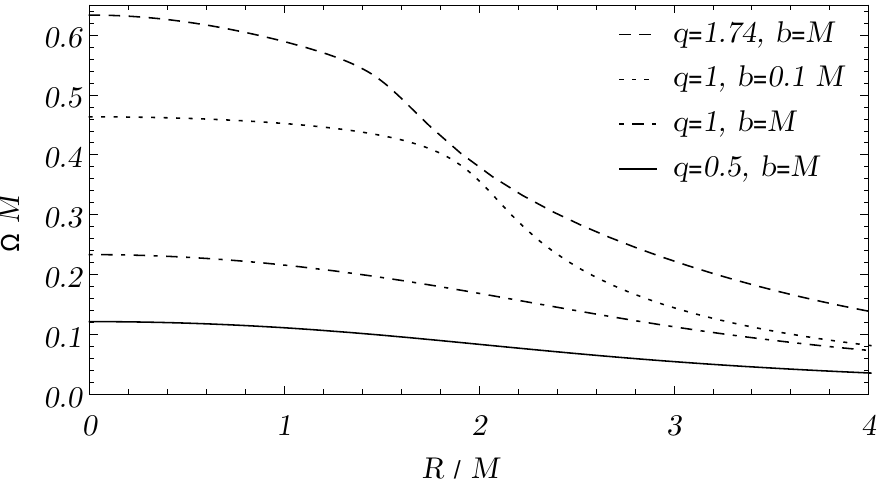}}
\caption{
Angular velocity of frames with the diagonal form of disk surface stress-energy tensor. 
The chosen values of parameters $q$ and $b$ imply disks which at some regions cannot be formed by CRGS.
\label{FIG_KerrOmega}
}
\end{figure}

\begin{figure}[b]
\centerline{\includegraphics[width=\linewidth]{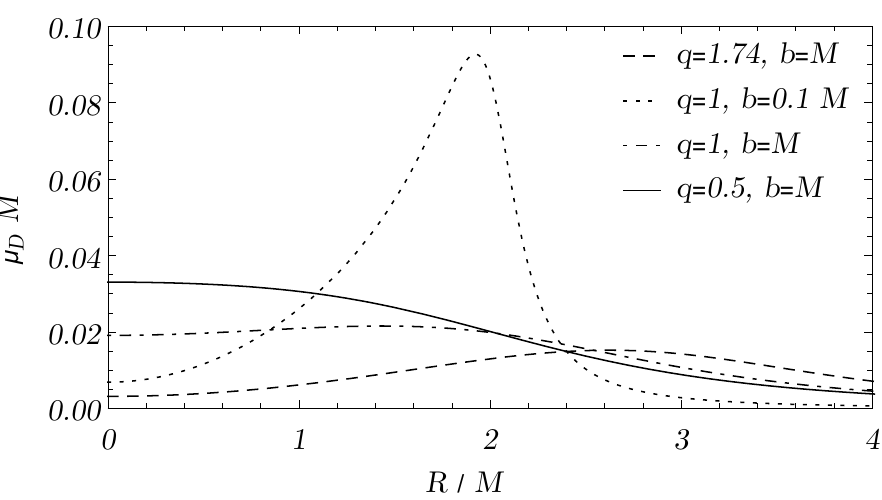}}
\centerline{\includegraphics[width=\linewidth]{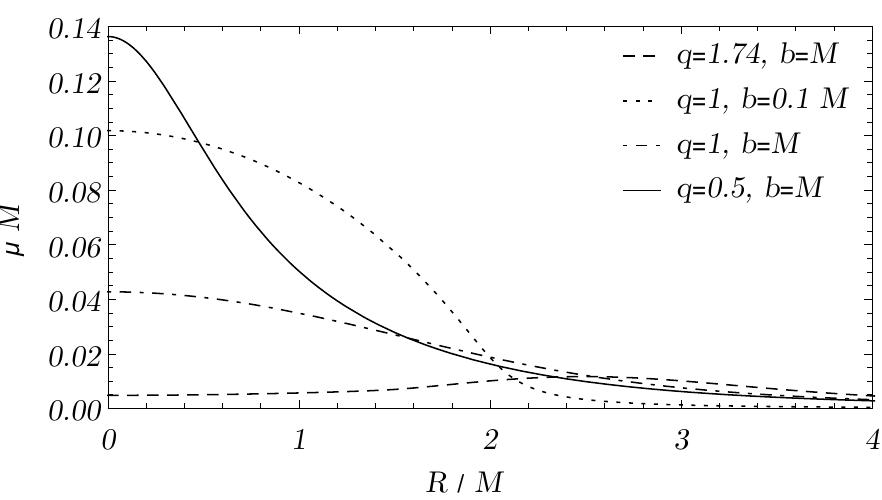}}
\centerline{\includegraphics[width=\linewidth]{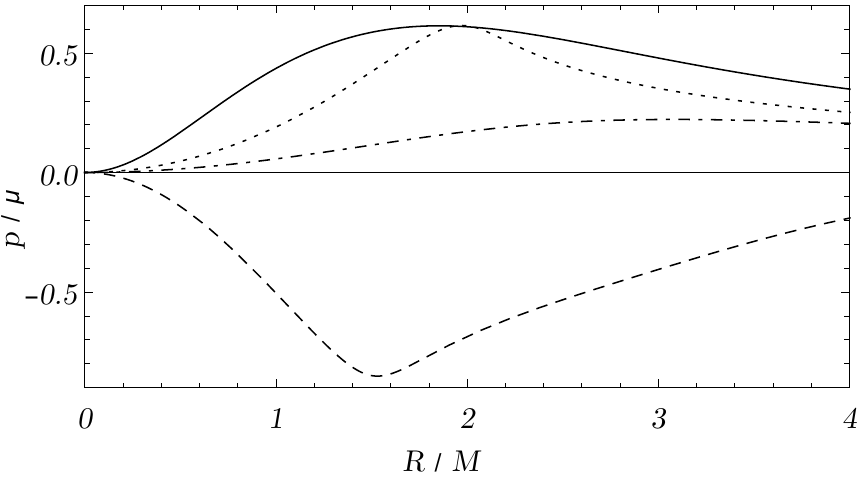}} 
\caption{
Komar mass density $\mu_D$ (top), stream density $\mu$ (middle) and ratio $p/\mu$ (bottom) for 
several Kerr disks.
\label{FIG_Kermumupe}
}
\end{figure}

In order to interpret the non-diagonal stress-energy tensor $S_{(a)(b)}$, we introduced observers \eqref{FIOtetrad}
with respect to whom the stress-tensor becomes diagonal (such observers are called FIOs, ``$\phi$-isotropic'', in \cite{BLPRL93}). In Fig.  \ref{FIG_KerrOmega} we show their
angular velocity \eqref{Omega_FIO} for several disk parameters. For a compact enough disk with $q>1$ we find
that the angular velocity slightly surpasses that of the horizon of an extreme Kerr black hole. However, if the
stress-energy tensor can be diagonalized we do not reach $\Omega \gtrsim 0.69/M$.
This angular velocity determines the speed of rotation of the RR disk source the properties of which are 
shown in Fig. \ref{FIG_Kermumupe}. Since the parameters were chosen to describe strong-field disk sources
we can see considerable differences between $\mu$ and $\mu_D$; and we also find the disk with a large negative pressure
$p\approx - \mu$. 

\begin{figure}[!htb]
\centerline{\includegraphics[width=\linewidth]{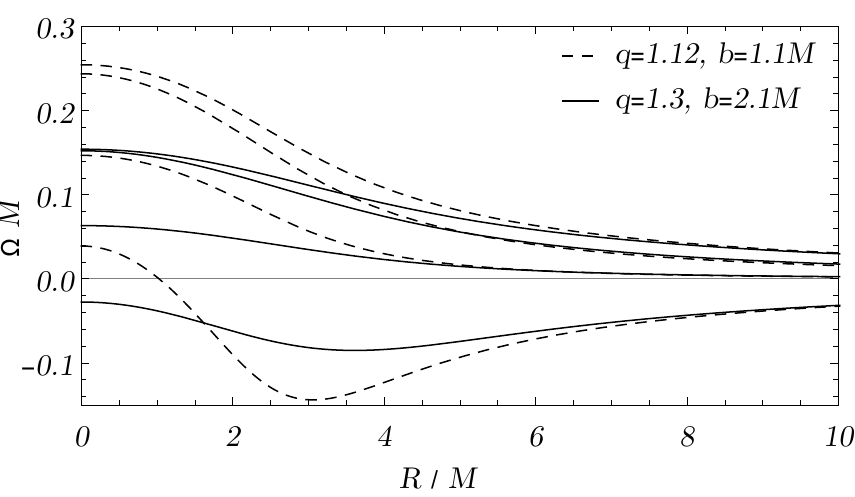}}
\caption{
Angular velocities (from top to bottom) of co-rotating geodesic stream $\Omega_+$, of 
isotropic observer $\Omega$, of zero angular momentum observer $\Omega_{\rm ZAMO}$, and of counter-rotating geodesic stream  $\Omega_-$ for two CRGS disk models.
\label{FIG_CGRSOmega}
}
\end{figure}

\begin{figure}[!htb]
\centerline{\includegraphics[width=\linewidth]{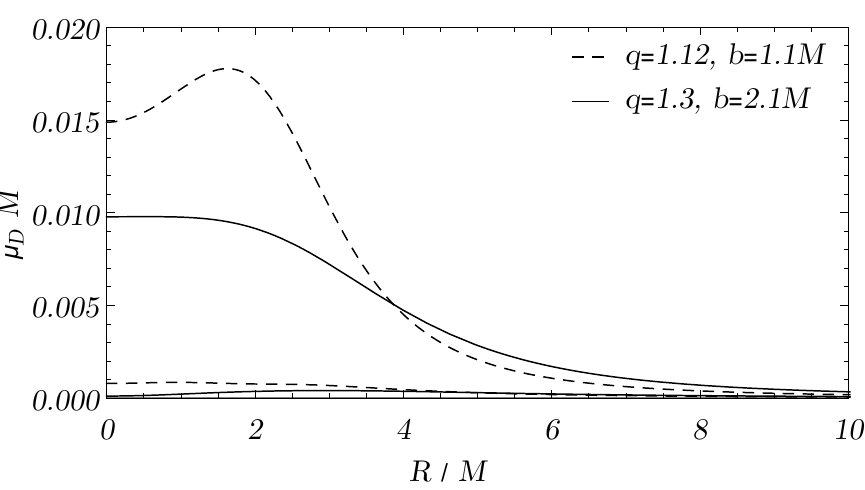}}
\caption{
Komar mass densities of co-rotating (upper curves) and 
counter-rotating (lower curves) streams.
Both sets of parameters $q$ and $b$ are chosen close to the  
limit when counter-rotating stream density becomes negative.
For chosen parameters the Komar mass of the counter-rotating components
is about $M/8$ only.
\label{FIG_CGRSMuD}
}
\end{figure}

The properties of the disks made of CRGS are shown in Fig. \ref{FIG_CGRSOmega}. For  two disk models with
quite different parameters $q$ and $b$ we illustrate the effects of the rotational dragging. It may force
both geodesic streams orbit to rotate
in the same direction w.r.t. infinity. Also, the stream densities differ significantly as is shown in  Fig.
\ref{FIG_CGRSMuD}.
For given mass and angular momentum of the disk, we can obtain disks of various compactness by varying parameter $b$. For larger values of $b$ dragging soon disappears, stream velocities 
become symmetric, $\Omega_- \approx -\Omega_+$, and the net angular momenta of  disks are 
due to the difference between $\mu_+$ and $\mu_-$.

\subsection*{Extreme Kerr disks}
The extreme Kerr black hole spacetime with $q=1$ is worth of a separate discussion. Many expressions 
simplify,
the condition \eqref{DefDeltaWEC} is always satisfied and for the RR model we can write explicitly
the energy density, pressure and velocity with respect to the ZAMO \eqref{basisZAMO} as follows:
\begin{align}
 \mu &= \frac{y}{4 \pi  M} \frac{w+x+2 y^2}{\left[(x+1)^2+y^2\right]^{3/2}},
   \\
 p_{\phi'}&=\frac{y}{4 \pi  M} \frac{w-x-2 y^2}{\left[(x+1)^2+y^2\right]^{3/2}},
   \\
 v_{\rm ZAMO}&=\frac{\sqrt{1-y^2}}{w+x+2} \frac{2+x(2 x+4+w)}{2+x(2x+4-w)},
\end{align}
where $w=\sqrt{x (x+4)+y^2+3}$. 

\begin{figure}[!h]
\centerline{\includegraphics[width=\linewidth]{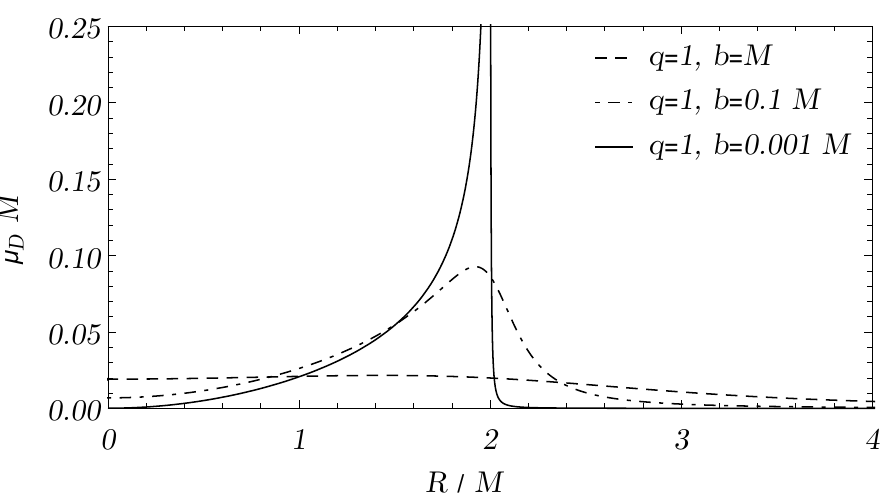}}
\centerline{\includegraphics[width=\linewidth]{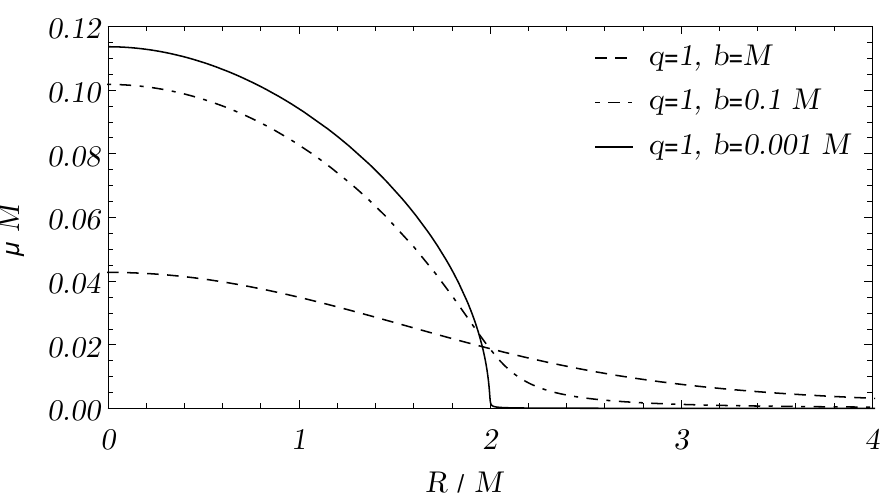}}
\centerline{\includegraphics[width=\linewidth]{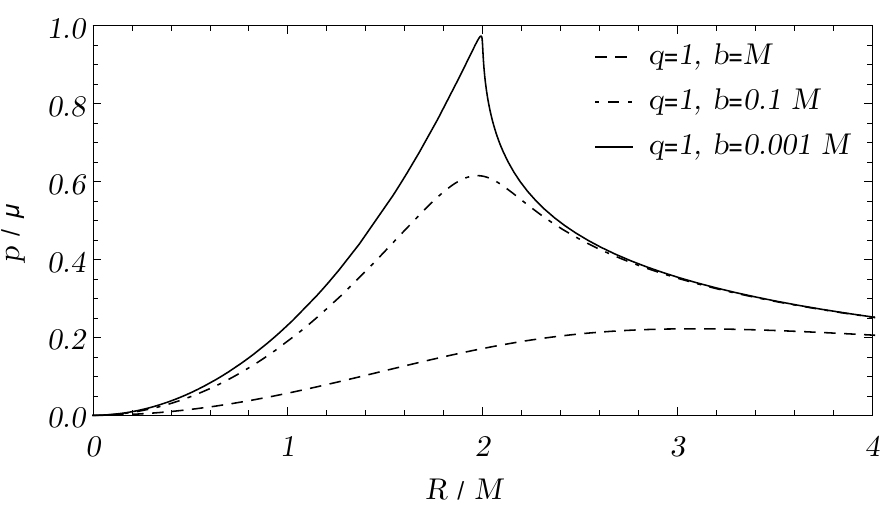}} 
\caption{
Komar mass density $\mu_D$ (top), RR stream density $\mu$ (middle) and ratio $p/\mu$ (bottom) for extreme Kerr disks for $b/M=1,0.1,0.001$. 
\label{FIG_q1}
}
\end{figure}

\begin{figure}[!h]
\centerline{\includegraphics[width=\linewidth]{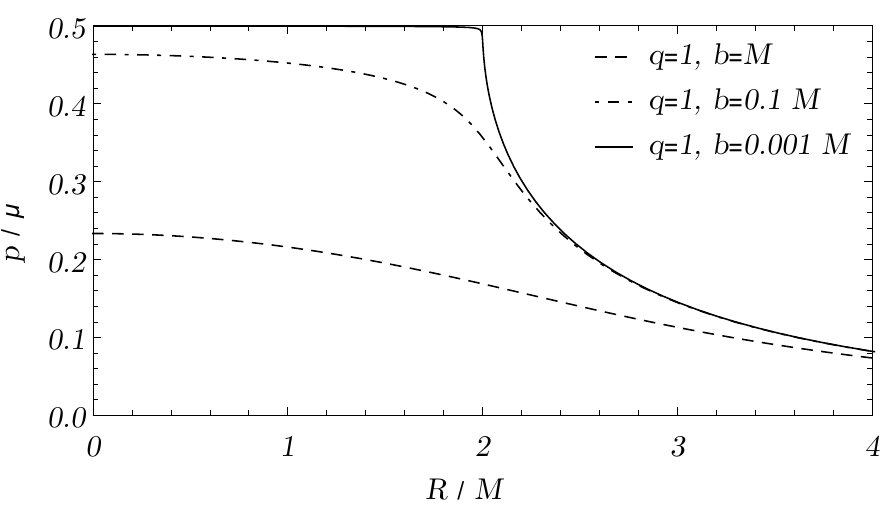}}
\caption{
Angular velocity with respect to infinity of rotating-ring sources of extreme Kerr disks. 
When $b\rightarrow 0$ all of the disks' mass is located below $R=2M$ and the rings exhibit rigid-body
rotation with angular velocity equal to the horizon angular velocity of an extreme Kerr black hole $M\Omega_H=1/2$.
\label{FIG_q1Omega}
}
\end{figure}

\begin{figure}[!h]
\centerline{\includegraphics[width=\linewidth]{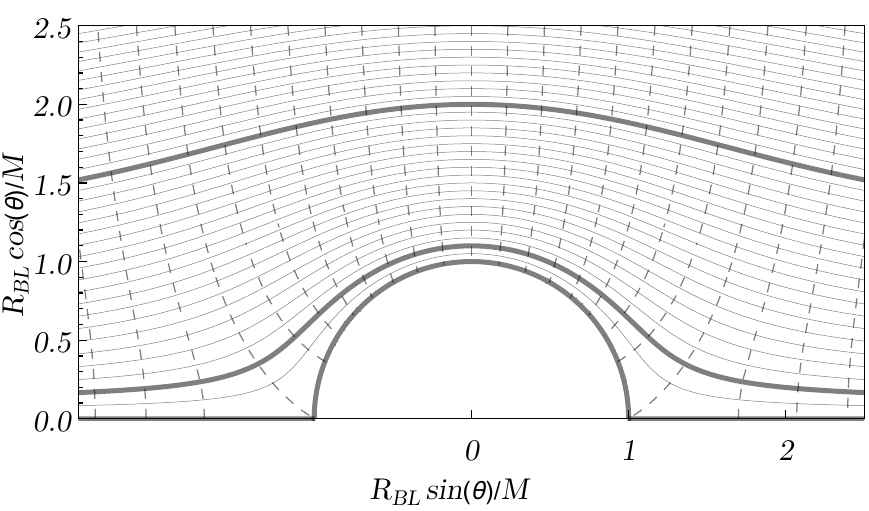}}
\caption{Shape of the $z=\rm const.$ identification surfaces in Boyer-Lindquist coordinates for extreme Kerr black hole. Thick lines are given by values $z/M=1,0.1,0.001$ considered in Fig. \ref{FIG_q1}. Dashed lines are places of constant circumferential radius $R/M=0,0.25,0.5, ...$.
\label{FIG_q1BLz}
}
\end{figure}

Also, only in the extreme Kerr case we can remove arbitrarily ``thin'' region between 
hypersurfaces ${}^-\Sigma$ and ${}^+\Sigma$, i.e., to consider the limit $b \rightarrow 0$, 
since for $0\le q < 1$ we would hit the horizon of the black hole and for $q>1$ the 
energy conditions would be violated for small $b$; and with parameter $\delta>1$ singularities would
appear outside the disk (see Fig. \ref{FIG_REGIO_KERR}). The RR disk with $q=1$ properties are illustrated in Fig. \ref{FIG_q1}. There we see that the limiting case is a finite-radius disk (cf. also Fig. \ref{FIG_KerrCummu}). 
Fig. \ref{FIG_q1Omega} shows that such disk rotates rigidly with angular velocity equal to the horizon angular velocity of an extreme Kerr black hole, $M\Omega_H=1/2$.
Similar behavior was observed in \cite{NM1993ApJ} for disks made of rigidly rotating dust.
In Fig. \ref{FIG_q1BLz} the shape of $z=\rm const.$  surfaces is shown in Boyer-Lindquist coordinates $R_{BL}, \theta$. One can check there that in the limiting case the extreme Kerr black hole 
is removed by the identification of mirrored points just above the extreme horizon 
($R=2M$ is the circumferential radius of the extreme horizon).

\subsection*{Tomimatsu-Sato $\delta=2$ disks}
The TS2 spacetime is determined by the mass $M$ and angular momentum
$M^2q$ \cite{Perjes89}.
The parameter $p$ again satisfies the relation $\kappa p^2 + q^2 = 1$,
$\kappa=\pm 1$. Introducing the abbreviations $a=x^2-\kappa$, $b=1-y^2$, 
$c=( \kappa p^2(x^4-1) - q^2(1-y^4) +2px\kappa a)$, 
and
$d=p^3xa(2(x^4\! -\! 1)+(\kappa x^2\!\! +\! 3)b)$, 
the line element is given by \eqref{WeyPap}, where
\begin{align}
e^{2 \nu}&=  {{p^4a^4+q^4b^4-2p^2q^2ab(2a^2+2b^2+3\kappa ab)}
                  \over
{c^2 + 4q^2y^2(  px\kappa a + (px+1)b  )^2}}~,\nonumber\\
 & & {}\nonumber\\
e^{2 \zeta}&= {{p^4a^4+q^4b^4-2p^2q^2ab(2a^2+2b^2+3\kappa ab)}
                                  \over
                                  {p^4(x^2-\kappa y^2)^4}}~,\\
 & & {}\nonumber\\
A&=  2 M q b 
  { -d\! -\! (4ax^2\!\! +\! (3\kappa x^2\!\! +\! 1)b)ap^2\!\! +\! q^2(px\! +\! 1)b^3
                  \over
                                  p^4a^4+q^4b^4-2p^2q^2ab(2a^2+2b^2+3\kappa ab)}~.\nonumber
\end{align}
These metric potentials
are written down in the short form which assumes $\kappa^2=1$ and thus it does not include the extreme ($q=1$) case,
which is identical with the extreme Kerr metric. This can be shown
by putting $\kappa=1, x=\hat{x}\delta/p,q=1$ and then taking limit $p\rightarrow 0$.

The static limit $q=0$ of the TS2 spacetime is the Darmois solution
(the Weyl metric with the metric potential $\nu$ proportional, with factor $2$,
to the metric potential of the Schwarzschild solution with mass $M/2$),
and the corresponding disk source are thus studied as
a special case of Zipoy-Vorhees metrics in \cite{BicakLyndKatz93}.

\begin{figure}[b]
\centerline{\includegraphics[width=8cm]{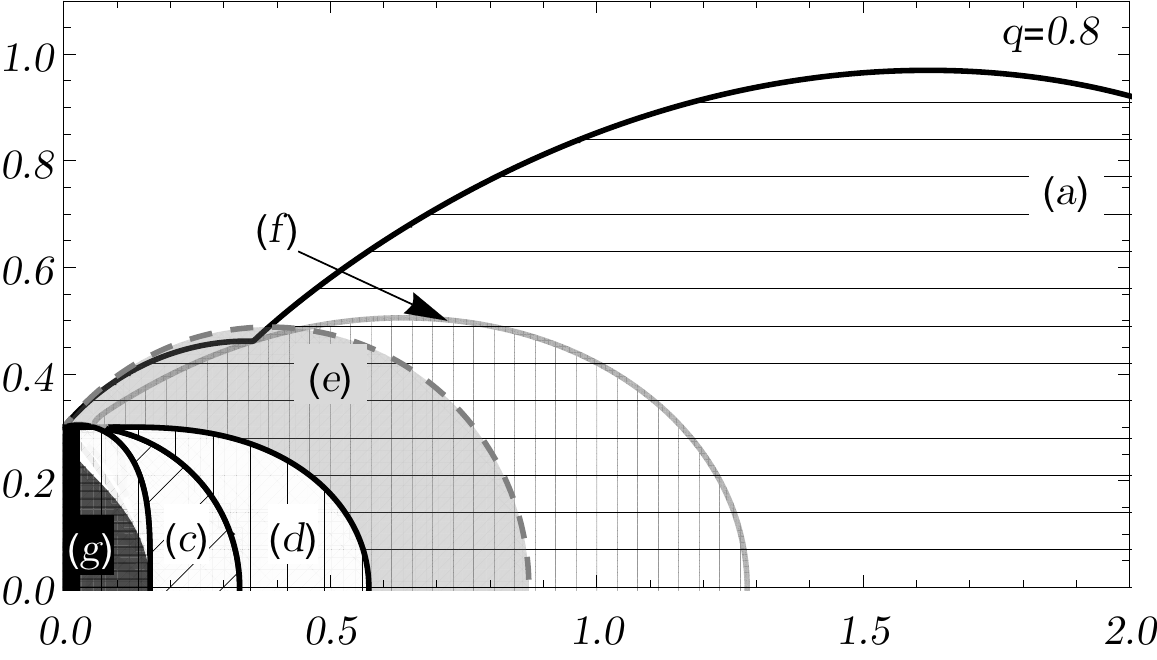}}
\centerline{\includegraphics[width=8cm]{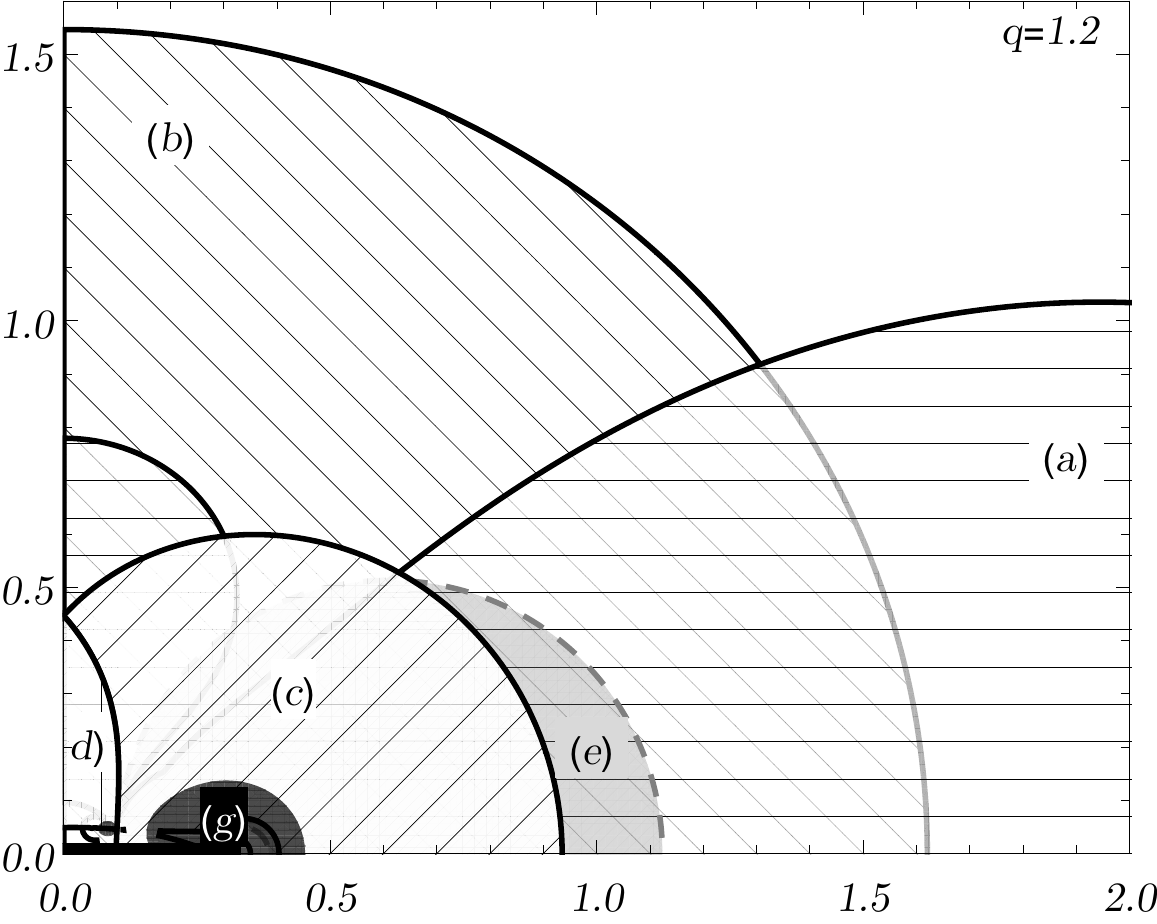}} 
\caption{
Properties of the stress-energy tensor of a $p_r=0$ disk as a source $\delta=2, M=1$ Tomimatsu-Sato geometry in the $\rho-z$ plane in the Weyl-Papapetrou coordinates for $q=0.8$ (top) and $q=1.2$ (bottom). When the identification surfaces $\Sigma^{\pm}$ are fixed by the prescription $z=\rm const.$ (which would appear as two horizontal lines in the $\rho-z$ plane) the plots show properties of the stress-energy tensor at given radius of the disk:
(a) one or both Keplerian velocities are superluminal;
(b) negative value of ${\rm det} S_{AB} \sim \mu p_{\phi'} \sim \mu_+ \mu_-$;
(c) no diagonalisation is possible, $\sigma<0$;
(d) energy density $\mu$ of RR is negative;
(e) ergosphere;
(f) pressure exceeding DEC limit, $p/\mu>1$;
(g) closed timelike curves.
\label{FIG_TS2}
}
\end{figure}

As mentioned in Section V, one of the most important differences between the Kerr and the TS2 family is that in the TS2 spacetime for $|q|>1$ there remain both singular rings and toroidal regions of closed timelike curves outside the equatorial plane. We limit the choice of the identification hypersurface in such a way, that singularities remain in the removed region.

In Fig. \ref{FIG_TS2} we illustrate the properties of the  stress-energy tensor of the TS $\delta=2$ disks for two values of parameter $q$, namely $q=0.8$ and $q=1.2$. For $|q|<1$ we observe DEC violation in region (f) as we approach the poles of the horizon. For $|q|>1$ we see how for smaller values RR model is not admissible because at some radii we cannot diagonalize the disk stress-energy tensor; this also indicates WEC violation. In both $|q|<1$ and $|q|>1$ cases we also see how incomplete set of timelike circular geodesics in the plane of the disk, region (a) disfavors CRGS model, which is, for $|q|>1$ even more strictly limited by the region of negative pressure (b), which, due to \eqref{detSAB}, implies one of the counter-rotating geodesic streams to have negative density.
These properties are  summarized for $0<q<2$ in the middle panel of Fig. \ref{FIG_REGIO_KERR}
where the dependence of the disk properties on the rotation parameter is illustrated.

\subsection*{Tomimatsu-Sato $\delta=3$ disks}
\begin{figure}[t]
\centerline{\includegraphics[width=44mm]{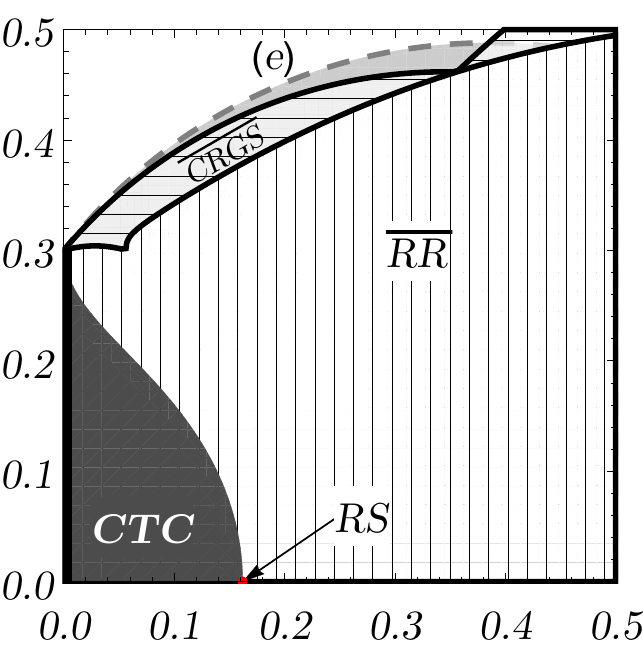}\includegraphics[width=44mm]{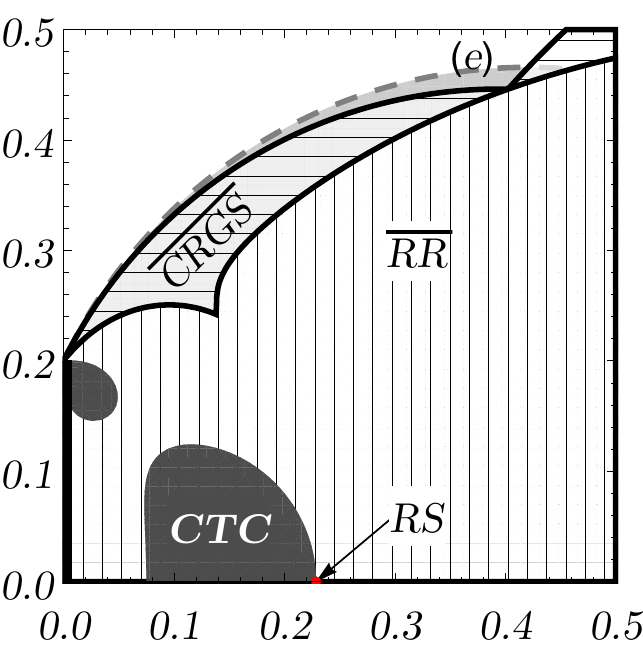}}
\caption{
Properties of disk sources with $\delta=2$ (left) and $\delta=3$ (right) for $q=0.8$ in a detailed view.
Figures show the central, strong field region in the $\rho-z$ plane in the Weyl-Papapetrou coordinates, $M=1$. 
All reasons why the disks cannot be made of particular matter model are considered and  
only combined regions where RR or CRGS models are not admissible are shown. From other features of the spacetime, 
only ergosphere (e) and the region of closed time-like curves (dark gray) is shown.
Arrows indicate position of ring singularities.
\label{FIG_TS32_08}
}
\end{figure}

\begin{figure}[!h]
\centerline{\includegraphics[width=44mm]{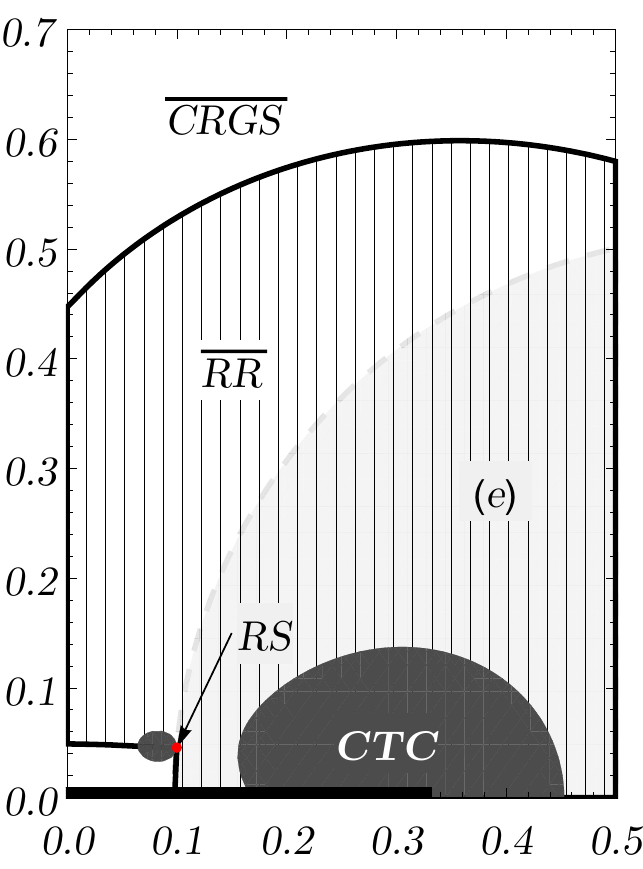}\includegraphics[width=44mm]{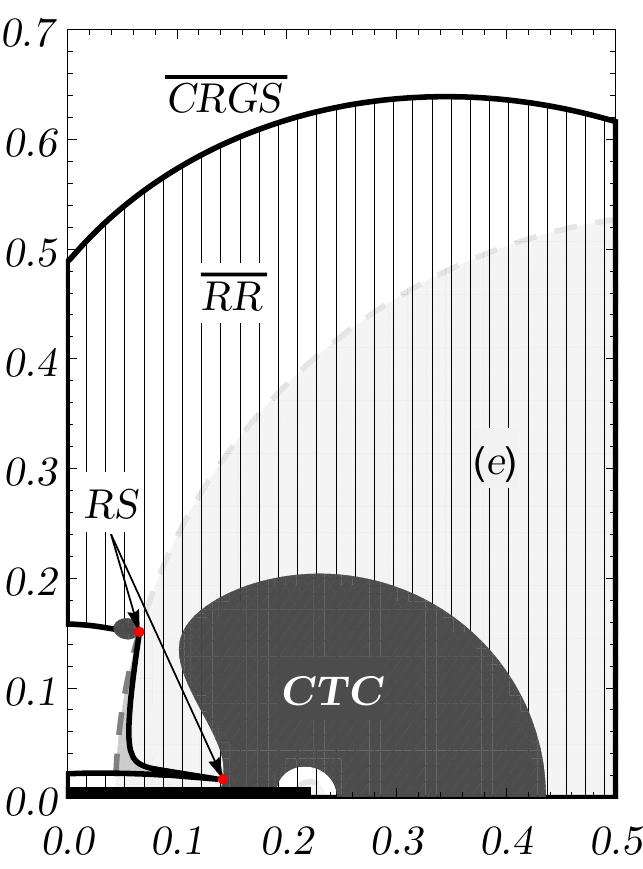}}
\caption{
Properties of disk sources with $\delta=2$ (left) and $\delta=3$ (right) for $q=1.2$ in a detailed view. Figures show the central, strong field region in the $\rho-z$ plane in the Weyl-Papapetrou coordinates, $M=1$. In the displayed range of the coordinate $z$, the stress-energy tensor cannot be made of CGRS. Otherwise, the caption of Fig \ref{FIG_TS32_08} applies here too. 
\label{FIG_TS32_12}
}
\end{figure}

For $\delta=3$ the metric potentials are given by rather lengthy expressions \cite{TomimatsuSato}, however, 
even though the central region fields are thus more complicated, the existence of both RR and CRGS disks is restricted already in weaker-field regions where the properties of the disk surface stress-energy tensors for $\delta=2$ and $\delta=3$ are analogous. 
The regions of parameters $q$ and $b$  where RR or CRGS models are applicable (see Fig. \ref{FIG_REGIO_KERR}) are very similar. We compare side by side the disk features for $\delta=2,3$ in Fig. \ref{FIG_TS32_08} and Fig. \ref{FIG_TS32_12}. These cover the central regions of the $\rho-z$ plane in the Weyl-Papapetrou coordinates, outside of which 
properties of $\delta=3$ disks are very similar to those of $\delta=2$ disks shown in Fig \ref{FIG_TS2}.


\section*{Conclusions}
The properties of the sources of stationary vacuum \hyphenation{axi-sym-met-ric} axisymmetric spacetimes
were analyzed by considering material disks which arise from suitable identifications of two hypersurfaces. We showed that at given point of the disk source the only degrees of freedom are two principal curvatures of the surface of identification. Choosing these to be zero yields disks without radial pressure. Since disks have no interiors we get direct relation of the properties of the disk sources and metric potentials and their first derivatives. This enables us to identify regions where certain features of the disk stress energy tensor hold for a given spacetime. We chose the first three members of the Tomimatsu-Sato family. Our approach is thus an extension of an analysis of such spacetimes where metric and its derivatives are used to find the causal structure, to describe the geodesic motion and to identify the presence of singularities.

The method of the construction by suitable identification enables us to avoid the pathological ``central regions'' and obtain completely regular spacetimes outside of the disks. The discontinuities in the first derivatives of the metric then represent material sources with plausible physical properties. The resulting disks have infinite radii but most of their mass appears near their center which is similar to the disk models of galaxies. Each disk source is parametrized by the properties of the corresponding spacetime -- by mass, angular momentum, the TS family index $\delta$ and by the parameter determining how large portion of a spacetime is replaced by the disk source.

The symmetries of spacetimes admit two interpretations of the surface stress-energy tensor of the disk with vanishing radial pressure. When the surface stress-energy tensor satisfies weak energy condition, we can consider the model of the rings defying the gravitational forces by their tangential pressure. Each ring rotates with an appropriate angular velocity which is the implication of the frame dragging and is the source of angular momentum. A mechanism which generates this pressure is not considered. When both co-rotating and counter-rotating circular timelike geodesics exist in the plane of the disk, the stress-energy tensor  satisfies energy conditions and has positive tangential pressure; rather surprisingly,  both circular streams may be endowed with appropriate positive densities and the Einstein equations then guarantee that these two densities can provide all three non-zero components of the disk surface stress-energy tensor. Differences in stream densities and velocities appear as the source of the net angular momentum.

The results show that for a given Tomimatsu-Sato spacetime we can construct disks of both  models if we exclude large enough region containing pathologies such as horizons, branching surfaces, naked singularities and closed timelike curves. We can see the influence of centrifugal effects for disks with large enough angular momentum which do not permit to have highly compact disks formed of counter-rotating geodesic streams with angular momentum $J \gg M^2$. Also, we provide another example when the extreme Kerr spacetime is confirmed as a limiting case for very compact extremely rotating disks. Since the spacetimes studied in this work contain regions of closed timelike curves we may consider a parametric collapse of the disk into the state when a region with CTC would appear above the disk. We then observe that energy conditions are violated {\it first}, before CTC appear.

\section*{Acknowledgments}
TL and JB acknowledge the partial support from the Grant GA\v{C}R 17-13525S of the Czech Republic. 
During the final preparation of the paper, JB was a visiting fellow at the Institute of Astronomy in Cambridge and at the Albert-Einstein Institute, Golm.
Thanks for the kind hospitality to R. G. McMahon and L. Andersson, and for discussions to N. W. Evans, G. W. Gibbons and J. E. Pringle. 

\FloatBarrier

\section*{APPENDIX A: STATIC DISKS}
Some characteristic features of the sources of axisymmetric stationary spacetimes 
can already be observed in case of static spacetimes. 
While in the main text we use the geometric approach \cite{Israel66} to get the stress energy tensor, here we will directly look at the components of the Einstein equations in the Weyl coordinates in which the metric reads
\cite{Stephani2009}
\begin{equation}
 ds^2 = - e^{2\nu} dt^2 + e^{2\zeta-2\nu}(d\rho^2+dz^2) + e^{2\nu}  \rho^2 d\phi^2.
\end{equation}

Assume we have two solutions $\nu^{\pm},\zeta^{\pm}$ of the field equations
\begin{align}
\Delta \nu &= 0,\\
\zeta_z &= 2\rho \nu_\rho \nu_z,\label{zeta_z}\\
\zeta_\rho &= \rho(\nu_\rho^2-\nu_z^2) \label{zeta_rho}
\end{align}
in regions $z>0$ and $z<0$ which are continuous across $z=0$, i.e., 
$\nu^+(\rho,z=0)=\nu^-(\rho,z=0)$ and $\zeta^+(\rho,z=0)=\zeta^-(\rho,z=0)$. 
Hence, also the radial partial derivatives of the potentials are equal along the plane $z=0$ and
\eqref{zeta_z} and \eqref{zeta_rho} guarantee that we either have 
continuity of normal derivatives of both potentials (hence no surface source) or the situation is
reflection-symmetric when
$\nu_z^+(\rho,z=0)=-\nu_z^-(\rho,z=0)$ and $\zeta_z^+(\rho,z=0)=-\nu_z^-(\rho,z=0)$.
Then the Einstein equations 
have two non-vanishing components which, under these assumptions, can be written as follows:
\begin{align}
G_{tt} = 2\,{e^{4\nu  -2\zeta  }} \left(   \rho\nu_\rho-1 \right)~ \Delta \nu
=8\pi S_{tt}~ \delta(z),\label{Gttstatic}
\\
G_{\phi\phi} = -2\,{e^{-2\zeta  }} {\rho}^{3} \nu_\rho ~ \Delta \nu 
=8\pi S_{\phi\phi}~ \delta(z).\label{Gffstatic}
\end{align}
Here $\Delta \nu = 2 \left. \nu^+_z\right|_{z=0} \delta(z)$, so these equations directly describe
how the discontinuity of the normal derivatives generates the components of the surface stress-energy tensor
$S_{tt}$ and  $S_{\phi\phi}$.  The remaining components of the Einstein tensor contain only the first derivatives 
of the potentials and vanish identically.

The reflection symmetry across $z=0$ can easily be achieved by using 
$\nu^-(\rho,z) = \nu^+(\rho,-z)$. Thus any solution of the Poisson
equation
$ 
 \Delta \nu = \lambda(\rho,z)
$ for any $\lambda$ such that $\lambda(\rho,z>0)=0$ can be used to generate 
a vacuum static spacetime with an axisymmetric disk source.

Therefore, we conclude that (i) gluing along the ``planes'' with the Weyl coordinate $z=\rm const.$ yields
no radial stresses,
(ii) the model of the surface stress-energy tensor has to provide a 
mechanism generating $S_{\phi\phi}$ as a consequence of which the disk matter resists the radial
acceleration, and (iii) while in the Newtonian limit we get $\Delta \nu = -4\pi \sigma \delta(z)$, 
(iv) in the strong field the sign of $-\Delta \nu$ and that of the energy density may not coincide
(e.g. due to the term $ \rho\nu_\rho-1$ in \eqref{Gttstatic}) so that the source cannot be then made from a realistic matter.

\section*{APPENDIX B: THE TOTAL MASS AND ANGULAR MOMENTUM}

The method of the construction of the disk sources we employed guarantees
that the values of the total mass and angular momentum given by an asymptotic behavior of the metric
are equal to those of the original
spacetimes. Using the Komar approach, the stationary spacetimes with Killing vectors 
corresponding to the time and axial symmetries enable us to write these quantities as integrals
of surface densities over the disk.

The mass $M$ defined by the asymptotic behavior of the metric can be expressed
as an integral of stress-energy tensor components
over spacelike hypersurface $\Sigma$ surrounding the regions of non-vanishing $T_{\mu\nu}$.

The Komar mass integral (see e.g. \cite{Wald84}),
\begin{equation} \label{Komar1}
M = -{1\over 8 \pi}\int\limits_{\partial\Sigma} \nabla^\mu\xi^\nu d\Sigma_{\mu \nu}~,
\end{equation}
can be evaluated over any 2-dimensional boundary of space-like hypersurface $\Sigma$
surrounding all sources $T^{\mu\nu}\ne 0$ so for disk-like sources we can integrate over the
area of the disk. For an antisymmetric tensor $A^{\mu\nu} $ the integration over upper part of the disk
takes the form
\begin{equation}
\int A^{\mu\nu} d\Sigma_{\mu\nu} = 
2 \int \left(A^{t\rho} z_{,s} - A^{tz} \rho_{,s}  \right) \;\sqrt{-g}\; ds\; d\phi.
\end{equation}
Since $A^{t\rho} z_{,s} - A^{tz} \rho_{,s} = A^{\mu\nu} \delta^{(t)}_\mu n_\nu \; e^{2\nu-2\zeta}\lambda^{-1}$, the timelike Killing vector $\xi^\mu = e^\nu e^\mu_{(t)}$ and
$\delta^{(t)}_\mu = - e^{-\nu}(e_{\mu(t)}+\rho^{-1} A e^{2\nu} e_{\mu(\phi)})$,
we can, regarding \eqref{KabDEF}, express the Komar mass (\ref{Komar1}) 
in terms of the discontinuity  of the first derivatives (denoted again as $[...]$)
\begin{align} 
\label{KomarMD} M_D &=  
-{1\over 4 \pi}\int_D\left( \left[ K_{(t)(t)}\right] + \frac{A e^{2\nu}}{\rho} \left[ K_{(t)(\phi)}\right] \right) \;dS\;,
\end{align}
where $dS = ~\rho\; \lambda^{-1}(s) ~ds ~d\phi$ is an area element inside the disk. 
This then yields Eq. \eqref{KomarMDX},
and if disks without the radial pressure are considered and the line element (\ref{WeyPap}) giving $g = -\rho^2~g_{\rho\rho}~g_{zz}$
is used, we get
\begin{equation}
\label{KomarMD2} M_D = \int_D{(S^\phi_\phi-S_t^t) ~\sqrt{g_{\rho\rho}}~2\pi\rho~d\rho}.
\end{equation}
Since we use the circumferential radius as a common  coordinate to compare the mass density curves of the disks with different parameters and with different intrinsic geometries, we transform this integral into the form \eqref{def_muD}.

The total angular momentum in an axisymmetric stationary spacetime can be defined by the expression \cite{Wald84}
\begin{equation} \label{KomarLikeJ1}
J = {1\over 16 \pi}\int\limits_{\partial\Sigma} \nabla^\mu\eta^\nu d\Sigma_{\mu \nu}~,
\end{equation}
where $\eta^\nu$ is the axial Killing vector field. 
Again, these integrals can be rewritten
as the integrals over the surface of the disk
\begin{align} 
\label{KomarJD} J_D =&  
-{1\over 8 \pi}\!\int_{\!D}\!\!
\Big\{ \left[ 
K_{(t)(\phi)}\right]\left(1+{\rho^{-2}}{A^2 e^{4\nu}}\right) 
\\\nonumber
&+ {\rho^{-1}}{A e^{2\nu}} \left(\left[ K_{(t)(t)}\right]+
\left[ K_{(\phi)(\phi)}\right]
\right) \Big\} \rho\, dS\;.
\end{align}
In terms of the metric potentials this turns into expression \eqref{KomarJDX} and for disks without a radial pressure we get  \eqref{def_jD}.
Apart from indicating the properties of the disk, the formulae for $M_D$ and $J_D$ can also be used as the independent checks of the code that evaluates $S_{ab}$, since the total spacetime mass and angular momentum are known. 

\bibliography{disks}

\begin{thebibliography}{42}%
\makeatletter
\providecommand \@ifxundefined [1]{%
 \@ifx{#1\undefined}
}%
\providecommand \@ifnum [1]{%
 \ifnum #1\expandafter \@firstoftwo
 \else \expandafter \@secondoftwo
 \fi
}%
\providecommand \@ifx [1]{%
 \ifx #1\expandafter \@firstoftwo
 \else \expandafter \@secondoftwo
 \fi
}%
\providecommand \natexlab [1]{#1}%
\providecommand \enquote  [1]{``#1''}%
\providecommand \bibnamefont  [1]{#1}%
\providecommand \bibfnamefont [1]{#1}%
\providecommand \citenamefont [1]{#1}%
\providecommand \href@noop [0]{\@secondoftwo}%
\providecommand \href [0]{\begingroup \@sanitize@url \@href}%
\providecommand \@href[1]{\@@startlink{#1}\@@href}%
\providecommand \@@href[1]{\endgroup#1\@@endlink}%
\providecommand \@sanitize@url [0]{\catcode `\\12\catcode `\$12\catcode
  `\&12\catcode `\#12\catcode `\^12\catcode `\_12\catcode `\%12\relax}%
\providecommand \@@startlink[1]{}%
\providecommand \@@endlink[0]{}%
\providecommand \url  [0]{\begingroup\@sanitize@url \@url }%
\providecommand \@url [1]{\endgroup\@href {#1}{\urlprefix }}%
\providecommand \urlprefix  [0]{URL }%
\providecommand \Eprint [0]{\href }%
\providecommand \doibase [0]{http://dx.doi.org/}%
\providecommand \selectlanguage [0]{\@gobble}%
\providecommand \bibinfo  [0]{\@secondoftwo}%
\providecommand \bibfield  [0]{\@secondoftwo}%
\providecommand \translation [1]{[#1]}%
\providecommand \BibitemOpen [0]{}%
\providecommand \bibitemStop [0]{}%
\providecommand \bibitemNoStop [0]{.\EOS\space}%
\providecommand \EOS [0]{\spacefactor3000\relax}%
\providecommand \BibitemShut  [1]{\csname bibitem#1\endcsname}%
\let\auto@bib@innerbib\@empty
%</preamble>
\bibitem [{\citenamefont {{Bi\v c\' ak}}\ \emph
  {et~al.}(1993{\natexlab{a}})\citenamefont {{Bi\v c\' ak}}, \citenamefont
  {{Lynden-Bell}},\ and\ \citenamefont {{Pichon}}}]{BicakLyndPichon93}%
  \BibitemOpen
  \bibfield  {author} {\bibinfo {author} {\bibfnamefont {J.}~\bibnamefont
  {{Bi\v c\' ak}}}, \bibinfo {author} {\bibfnamefont {D.}~\bibnamefont
  {{Lynden-Bell}}}, \ and\ \bibinfo {author} {\bibfnamefont {C.}~\bibnamefont
  {{Pichon}}},\ }\bibfield  {title} {\enquote {\bibinfo {title} {{Relativistic
  Discs and Flat Galaxy Models}},}\ }\href {\doibase 10.1093/mnras/265.1.126}
  {\bibfield  {journal} {\bibinfo  {journal} {Mon. Not. R. Astron. Soc.}\
  }\textbf {\bibinfo {volume} {265}},\ \bibinfo {pages} {126--144} (\bibinfo
  {year} {1993}{\natexlab{a}})}\BibitemShut {NoStop}%
\bibitem [{\citenamefont {{Evans}}\ and\ \citenamefont {{de
  Zeeuw}}(1992)}]{Evans}%
  \BibitemOpen
  \bibfield  {author} {\bibinfo {author} {\bibfnamefont {N.~W.}\ \bibnamefont
  {{Evans}}}\ and\ \bibinfo {author} {\bibfnamefont {P.~T.}\ \bibnamefont {{de
  Zeeuw}}},\ }\bibfield  {title} {\enquote {\bibinfo {title}
  {{Potential-density pairs for flat galaxies}},}\ }\href {\doibase
  10.1093/mnras/257.1.152} {\bibfield  {journal} {\bibinfo  {journal} {Mon.
  Not. R. Astron. Soc.}\ }\textbf {\bibinfo {volume} {257}},\ \bibinfo {pages}
  {152--176} (\bibinfo {year} {1992})}\BibitemShut {NoStop}%
\bibitem [{\citenamefont {{Bi\v c\' ak}}\ \emph
  {et~al.}(1993{\natexlab{b}})\citenamefont {{Bi\v c\' ak}}, \citenamefont
  {{Lynden-Bell}},\ and\ \citenamefont {{Katz}}}]{BicakLyndKatz93}%
  \BibitemOpen
  \bibfield  {author} {\bibinfo {author} {\bibfnamefont {J.}~\bibnamefont
  {{Bi\v c\' ak}}}, \bibinfo {author} {\bibfnamefont {D.}~\bibnamefont
  {{Lynden-Bell}}}, \ and\ \bibinfo {author} {\bibfnamefont {J.}~\bibnamefont
  {{Katz}}},\ }\bibfield  {title} {\enquote {\bibinfo {title} {{Relativistic
  disks as sources of static vacuum spacetimes}},}\ }\href {\doibase
  10.1103/PhysRevD.47.4334} {\bibfield  {journal} {\bibinfo  {journal} {\prd}\
  }\textbf {\bibinfo {volume} {47}},\ \bibinfo {pages} {4334--4343} (\bibinfo
  {year} {1993}{\natexlab{b}})}\BibitemShut {NoStop}%
\bibitem [{\citenamefont {{Neugebauer}}\ and\ \citenamefont
  {{Meinel}}(1995)}]{Neugebauer1995}%
  \BibitemOpen
  \bibfield  {author} {\bibinfo {author} {\bibfnamefont {G.}~\bibnamefont
  {{Neugebauer}}}\ and\ \bibinfo {author} {\bibfnamefont {R.}~\bibnamefont
  {{Meinel}}},\ }\bibfield  {title} {\enquote {\bibinfo {title} {{General
  Relativistic Gravitational Field of a Rigidly Rotating Disk of Dust: Solution
  in Terms of Ultraelliptic Functions}},}\ }\href {\doibase
  10.1103/PhysRevLett.75.3046} {\bibfield  {journal} {\bibinfo  {journal}
  {Phys. Rev. Lett.}\ }\textbf {\bibinfo {volume} {75}},\ \bibinfo {pages}
  {3046--3047} (\bibinfo {year} {1995})},\ \Eprint
  {http://arxiv.org/abs/gr-qc/0302060} {gr-qc/0302060} \BibitemShut {NoStop}%
\bibitem [{\citenamefont {{Neugebauer}}\ \emph {et~al.}(1996)\citenamefont
  {{Neugebauer}}, \citenamefont {{Kleinwaechter}},\ and\ \citenamefont
  {{Meinel}}}]{Neugebauer1996}%
  \BibitemOpen
  \bibfield  {author} {\bibinfo {author} {\bibfnamefont {G.}~\bibnamefont
  {{Neugebauer}}}, \bibinfo {author} {\bibfnamefont {A.}~\bibnamefont
  {{Kleinwaechter}}}, \ and\ \bibinfo {author} {\bibfnamefont {R.}~\bibnamefont
  {{Meinel}}},\ }\bibfield  {title} {\enquote {\bibinfo {title}
  {{Relativistically rotating dust.}}}\ }\href@noop {} {\bibfield  {journal}
  {\bibinfo  {journal} {Helvetica Physica Acta}\ }\textbf {\bibinfo {volume}
  {69}},\ \bibinfo {pages} {472--489} (\bibinfo {year} {1996})},\ \Eprint
  {http://arxiv.org/abs/gr-qc/0301107} {gr-qc/0301107} \BibitemShut {NoStop}%
\bibitem [{\citenamefont {{Klein}}\ and\ \citenamefont
  {{Richter}}(1999)}]{Klein1999}%
  \BibitemOpen
  \bibfield  {author} {\bibinfo {author} {\bibfnamefont {C.}~\bibnamefont
  {{Klein}}}\ and\ \bibinfo {author} {\bibfnamefont {O.}~\bibnamefont
  {{Richter}}},\ }\bibfield  {title} {\enquote {\bibinfo {title} {{Exact
  Relativistic Gravitational Field of a Stationary Counterrotating Dust
  Disk}},}\ }\href {\doibase 10.1103/PhysRevLett.83.2884} {\bibfield  {journal}
  {\bibinfo  {journal} {Phys. Rev. Lett.}\ }\textbf {\bibinfo {volume} {83}},\
  \bibinfo {pages} {2884--2887} (\bibinfo {year} {1999})},\ \Eprint
  {http://arxiv.org/abs/gr-qc/9908042} {gr-qc/9908042} \BibitemShut {NoStop}%
\bibitem [{\citenamefont {{Frauendiener}}\ and\ \citenamefont
  {{Klein}}(2001)}]{Klein2001}%
  \BibitemOpen
  \bibfield  {author} {\bibinfo {author} {\bibfnamefont {J.}~\bibnamefont
  {{Frauendiener}}}\ and\ \bibinfo {author} {\bibfnamefont {C.}~\bibnamefont
  {{Klein}}},\ }\bibfield  {title} {\enquote {\bibinfo {title} {{Exact
  relativistic treatment of stationary counterrotating dust disks: Physical
  properties}},}\ }\href {\doibase 10.1103/PhysRevD.63.084025} {\bibfield
  {journal} {\bibinfo  {journal} {\prd}\ }\textbf {\bibinfo {volume} {63}},\
  \bibinfo {eid} {084025} (\bibinfo {year} {2001})},\ \Eprint
  {http://arxiv.org/abs/gr-qc/0102096} {gr-qc/0102096} \BibitemShut {NoStop}%
\bibitem [{\citenamefont {{Morgan}}\ and\ \citenamefont
  {{Morgan}}(1969)}]{MorganMorgan}%
  \BibitemOpen
  \bibfield  {author} {\bibinfo {author} {\bibfnamefont {T.}~\bibnamefont
  {{Morgan}}}\ and\ \bibinfo {author} {\bibfnamefont {L.}~\bibnamefont
  {{Morgan}}},\ }\bibfield  {title} {\enquote {\bibinfo {title} {{The
  Gravitational Field of a Disk}},}\ }\href {\doibase 10.1103/PhysRev.183.1097}
  {\bibfield  {journal} {\bibinfo  {journal} {Phys. Rev.}\ }\textbf {\bibinfo
  {volume} {183}},\ \bibinfo {pages} {1097--1101} (\bibinfo {year}
  {1969})}\BibitemShut {NoStop}%
\bibitem [{\citenamefont {{Bi\v c\' ak}}\ and\ \citenamefont
  {{Ledvinka}}(1993)}]{BLPRL93}%
  \BibitemOpen
  \bibfield  {author} {\bibinfo {author} {\bibfnamefont {J.}~\bibnamefont
  {{Bi\v c\' ak}}}\ and\ \bibinfo {author} {\bibfnamefont {T.}~\bibnamefont
  {{Ledvinka}}},\ }\bibfield  {title} {\enquote {\bibinfo {title}
  {{Relativistic Disks as sources of the Kerr metric}},}\ }\href {\doibase
  10.1103/PhysRevLett.71.1669} {\bibfield  {journal} {\bibinfo  {journal}
  {Phys. Rev. Lett.}\ }\textbf {\bibinfo {volume} {71}},\ \bibinfo {pages}
  {1669--1672} (\bibinfo {year} {1993})}\BibitemShut {NoStop}%
\bibitem [{\citenamefont {{Pichon}}\ and\ \citenamefont
  {{Lynden-Bell}}(1996)}]{Pichon96}%
  \BibitemOpen
  \bibfield  {author} {\bibinfo {author} {\bibfnamefont {C.}~\bibnamefont
  {{Pichon}}}\ and\ \bibinfo {author} {\bibfnamefont {D.}~\bibnamefont
  {{Lynden-Bell}}},\ }\bibfield  {title} {\enquote {\bibinfo {title} {{New
  sources for Kerr and other metrics: rotating relativistic discs with pressure
  support}},}\ }\href {\doibase 10.1093/mnras/280.4.1007} {\bibfield  {journal}
  {\bibinfo  {journal} {Mon. Not. R. Astron. Soc.}\ }\textbf {\bibinfo {volume}
  {280}},\ \bibinfo {pages} {1007--1026} (\bibinfo {year} {1996})},\ \Eprint
  {http://arxiv.org/abs/astro-ph/9605037} {astro-ph/9605037} \BibitemShut
  {NoStop}%
\bibitem [{\citenamefont {{Vogt}}\ and\ \citenamefont
  {{Letelier}}(2005{\natexlab{a}})}]{VogtLet1}%
  \BibitemOpen
  \bibfield  {author} {\bibinfo {author} {\bibfnamefont {D.}~\bibnamefont
  {{Vogt}}}\ and\ \bibinfo {author} {\bibfnamefont {P.~S.}\ \bibnamefont
  {{Letelier}}},\ }\bibfield  {title} {\enquote {\bibinfo {title}
  {{Relativistic models of galaxies}},}\ }\href {\doibase
  10.1111/j.1365-2966.2005.09436.x} {\bibfield  {journal} {\bibinfo  {journal}
  {Mon. Not. R. Astron. Soc.}\ }\textbf {\bibinfo {volume} {363}},\ \bibinfo
  {pages} {268--284} (\bibinfo {year} {2005}{\natexlab{a}})},\ \Eprint
  {http://arxiv.org/abs/astro-ph/0507406} {astro-ph/0507406} \BibitemShut
  {NoStop}%
\bibitem [{\citenamefont {{Gonz{\'a}lez}}\ and\ \citenamefont
  {{Guti{\'e}rrez-Pi{\~n}eres}}(2012)}]{Gonzales2012}%
  \BibitemOpen
  \bibfield  {author} {\bibinfo {author} {\bibfnamefont {G.~A.}\ \bibnamefont
  {{Gonz{\'a}lez}}}\ and\ \bibinfo {author} {\bibfnamefont {A.~C.}\
  \bibnamefont {{Guti{\'e}rrez-Pi{\~n}eres}}},\ }\bibfield  {title} {\enquote
  {\bibinfo {title} {{Stationary axially symmetric relativistic thin discs with
  nonzero radial pressure}},}\ }\href {\doibase 10.1088/0264-9381/29/13/135001}
  {\bibfield  {journal} {\bibinfo  {journal} {Class. Quantum Grav.}\ }\textbf
  {\bibinfo {volume} {29}},\ \bibinfo {eid} {135001} (\bibinfo {year}
  {2012})}\BibitemShut {NoStop}%
\bibitem [{\citenamefont {{Guti{\'e}rrez-Pi{\~n}eres}}\ \emph
  {et~al.}(2013)\citenamefont {{Guti{\'e}rrez-Pi{\~n}eres}}, \citenamefont
  {{Gonz{\'a}lez}},\ and\ \citenamefont {{Quevedo}}}]{Antonio}%
  \BibitemOpen
  \bibfield  {author} {\bibinfo {author} {\bibfnamefont {A.~C.}\ \bibnamefont
  {{Guti{\'e}rrez-Pi{\~n}eres}}}, \bibinfo {author} {\bibfnamefont {G.~A.}\
  \bibnamefont {{Gonz{\'a}lez}}}, \ and\ \bibinfo {author} {\bibfnamefont
  {H.}~\bibnamefont {{Quevedo}}},\ }\bibfield  {title} {\enquote {\bibinfo
  {title} {{Conformastatic disk-haloes in Einstein-Maxwell gravity}},}\ }\href
  {\doibase 10.1103/PhysRevD.87.044010} {\bibfield  {journal} {\bibinfo
  {journal} {\prd}\ }\textbf {\bibinfo {volume} {87}},\ \bibinfo {eid} {044010}
  (\bibinfo {year} {2013})},\ \Eprint {http://arxiv.org/abs/1211.4941}
  {arXiv:1211.4941 [gr-qc]} \BibitemShut {NoStop}%
\bibitem [{\citenamefont {{Ledvinka}}\ \emph {et~al.}(1999)\citenamefont
  {{Ledvinka}}, \citenamefont {{\v Zofka}},\ and\ \citenamefont {{Bi\v c\'
  ak}}}]{Zofka}%
  \BibitemOpen
  \bibfield  {author} {\bibinfo {author} {\bibfnamefont {T.}~\bibnamefont
  {{Ledvinka}}}, \bibinfo {author} {\bibfnamefont {M.}~\bibnamefont {{\v
  Zofka}}}, \ and\ \bibinfo {author} {\bibfnamefont {J.}~\bibnamefont {{Bi\v
  c\' ak}}},\ }\bibfield  {title} {\enquote {\bibinfo {title} {{Relativistic
  Disks as Sources of Kerr-Newman Fields}},}\ }in\ \href@noop {} {\emph
  {\bibinfo {booktitle} {Recent Developments in Theoretical and Experimental
  General Relativity, Gravitation, and Relativistic Field Theories, Proceedings
  of MG8 Meeting, Jerusalem}}},\ \bibinfo {editor} {edited by\ \bibinfo
  {editor} {\bibfnamefont {T.}~\bibnamefont {{Piran}}}\ and\ \bibinfo {editor}
  {\bibfnamefont {R.}~\bibnamefont {{Ruffini}}}}\ (\bibinfo  {publisher} {World
  Scientific},\ \bibinfo {address} {Singapore},\ \bibinfo {year} {1999})\ p.\
  \bibinfo {pages} {339},\ \Eprint {http://arxiv.org/abs/gr-qc/9801053}
  {gr-qc/9801053} \BibitemShut {NoStop}%
\bibitem [{\citenamefont {{Klein}}(2003)}]{Klein2003}%
  \BibitemOpen
  \bibfield  {author} {\bibinfo {author} {\bibfnamefont {C.}~\bibnamefont
  {{Klein}}},\ }\bibfield  {title} {\enquote {\bibinfo {title} {{On explicit
  solutions to the stationary axisymmetric Einstein-Maxwell equations
  describing dust disks}},}\ }\href {\doibase 10.1002/andp.200310029}
  {\bibfield  {journal} {\bibinfo  {journal} {Annalen der Physik}\ }\textbf
  {\bibinfo {volume} {515}},\ \bibinfo {pages} {599--639} (\bibinfo {year}
  {2003})},\ \Eprint {http://arxiv.org/abs/gr-qc/0512134} {gr-qc/0512134}
  \BibitemShut {NoStop}%
\bibitem [{\citenamefont {{Lemos}}\ and\ \citenamefont
  {{Letelier}}(1994)}]{Lemos}%
  \BibitemOpen
  \bibfield  {author} {\bibinfo {author} {\bibfnamefont {J.~P.~S.}\
  \bibnamefont {{Lemos}}}\ and\ \bibinfo {author} {\bibfnamefont {P.~S.}\
  \bibnamefont {{Letelier}}},\ }\bibfield  {title} {\enquote {\bibinfo {title}
  {{Exact general relativistic thin disks around black holes}},}\ }\href
  {\doibase 10.1103/PhysRevD.49.5135} {\bibfield  {journal} {\bibinfo
  {journal} {\prd}\ }\textbf {\bibinfo {volume} {49}},\ \bibinfo {pages}
  {5135--5143} (\bibinfo {year} {1994})}\BibitemShut {NoStop}%
\bibitem [{\citenamefont {{Vogt}}\ and\ \citenamefont
  {{Letelier}}(2005{\natexlab{b}})}]{VogtLet2}%
  \BibitemOpen
  \bibfield  {author} {\bibinfo {author} {\bibfnamefont {D.}~\bibnamefont
  {{Vogt}}}\ and\ \bibinfo {author} {\bibfnamefont {P.~S.}\ \bibnamefont
  {{Letelier}}},\ }\bibfield  {title} {\enquote {\bibinfo {title} {{General
  relativistic model for the gravitational field of active galactic nuclei
  surrounded by a disk}},}\ }\href {\doibase 10.1103/PhysRevD.71.044009}
  {\bibfield  {journal} {\bibinfo  {journal} {\prd}\ }\textbf {\bibinfo
  {volume} {71}},\ \bibinfo {eid} {044009} (\bibinfo {year}
  {2005}{\natexlab{b}})},\ \Eprint {http://arxiv.org/abs/gr-qc/0409109}
  {gr-qc/0409109} \BibitemShut {NoStop}%
\bibitem [{\citenamefont {{Karas}}\ \emph {et~al.}(2004)\citenamefont
  {{Karas}}, \citenamefont {{Hur{\'e}}},\ and\ \citenamefont
  {{Semer{\'a}k}}}]{KarasHure}%
  \BibitemOpen
  \bibfield  {author} {\bibinfo {author} {\bibfnamefont {V.}~\bibnamefont
  {{Karas}}}, \bibinfo {author} {\bibfnamefont {J.-M.}\ \bibnamefont
  {{Hur{\'e}}}}, \ and\ \bibinfo {author} {\bibfnamefont {O.}~\bibnamefont
  {{Semer{\'a}k}}},\ }\bibfield  {title} {\enquote {\bibinfo {title} {{TOPICAL
  REVIEW: Gravitating discs around black holes}},}\ }\href {\doibase
  10.1088/0264-9381/21/7/R01} {\bibfield  {journal} {\bibinfo  {journal}
  {Class. Quantum Grav.}\ }\textbf {\bibinfo {volume} {21}},\ \bibinfo {pages}
  {R1--R51} (\bibinfo {year} {2004})},\ \Eprint
  {http://arxiv.org/abs/astro-ph/0401345} {astro-ph/0401345} \BibitemShut
  {NoStop}%
\bibitem [{\citenamefont {{Sukov{\'a}}}\ and\ \citenamefont
  {{Semer{\'a}k}}(2013)}]{Semerak}%
  \BibitemOpen
  \bibfield  {author} {\bibinfo {author} {\bibfnamefont {P.}~\bibnamefont
  {{Sukov{\'a}}}}\ and\ \bibinfo {author} {\bibfnamefont {O.}~\bibnamefont
  {{Semer{\'a}k}}},\ }\bibfield  {title} {\enquote {\bibinfo {title} {{Free
  motion around black holes with discs or rings: between integrability and
  chaos - III}},}\ }\href {\doibase 10.1093/mnras/stt1587} {\bibfield
  {journal} {\bibinfo  {journal} {Mon. Not. R. Astron. Soc.}\ }\textbf
  {\bibinfo {volume} {436}},\ \bibinfo {pages} {978--996} (\bibinfo {year}
  {2013})},\ \Eprint {http://arxiv.org/abs/1308.4306} {arXiv:1308.4306 [gr-qc]}
  \BibitemShut {NoStop}%
\bibitem [{\citenamefont {{Tomimatsu}}\ and\ \citenamefont
  {{Sato}}(1973{\natexlab{a}})}]{TomimatsuSato}%
  \BibitemOpen
  \bibfield  {author} {\bibinfo {author} {\bibfnamefont {A.}~\bibnamefont
  {{Tomimatsu}}}\ and\ \bibinfo {author} {\bibfnamefont {H.}~\bibnamefont
  {{Sato}}},\ }\bibfield  {title} {\enquote {\bibinfo {title} {{New Series of
  Exact Solutions for Gravitational Fields of Spinning Masses}},}\ }\href
  {\doibase 10.1143/PTP.50.95} {\bibfield  {journal} {\bibinfo  {journal}
  {Prog. Theor. Phys.}\ }\textbf {\bibinfo {volume} {50}},\ \bibinfo {pages}
  {95--110} (\bibinfo {year} {1973}{\natexlab{a}})}\BibitemShut {NoStop}%
\bibitem [{\citenamefont {{Tomimatsu}}\ and\ \citenamefont
  {{Sato}}(1973{\natexlab{b}})}]{TomimatsuSatoNC73}%
  \BibitemOpen
  \bibfield  {author} {\bibinfo {author} {\bibfnamefont {A.}~\bibnamefont
  {{Tomimatsu}}}\ and\ \bibinfo {author} {\bibfnamefont {H.}~\bibnamefont
  {{Sato}}},\ }\bibfield  {title} {\enquote {\bibinfo {title} {{Event horizon
  of the Tomimatsu-Sato metrics}},}\ }\href {\doibase 10.1007/BF02725358}
  {\bibfield  {journal} {\bibinfo  {journal} {Lett. Nuovo Cimento}\ }\textbf
  {\bibinfo {volume} {8}},\ \bibinfo {pages} {740--742} (\bibinfo {year}
  {1973}{\natexlab{b}})}\BibitemShut {NoStop}%
\bibitem [{\citenamefont {{Gibbons}}\ and\ \citenamefont
  {{Russell-Clark}}(1973)}]{Gibbons}%
  \BibitemOpen
  \bibfield  {author} {\bibinfo {author} {\bibfnamefont {G.~W.}\ \bibnamefont
  {{Gibbons}}}\ and\ \bibinfo {author} {\bibfnamefont {R.~A.}\ \bibnamefont
  {{Russell-Clark}}},\ }\bibfield  {title} {\enquote {\bibinfo {title} {{Note
  on the Sato-Tomimatsu Solution of Einstein's Equations}},}\ }\href {\doibase
  10.1103/PhysRevLett.30.398} {\bibfield  {journal} {\bibinfo  {journal} {Phys.
  Rev. Lett.}\ }\textbf {\bibinfo {volume} {30}},\ \bibinfo {pages} {398--399}
  (\bibinfo {year} {1973})}\BibitemShut {NoStop}%
\bibitem [{\citenamefont {{Kodama}}\ and\ \citenamefont
  {{Hikida}}(2003)}]{Kodama}%
  \BibitemOpen
  \bibfield  {author} {\bibinfo {author} {\bibfnamefont {H.}~\bibnamefont
  {{Kodama}}}\ and\ \bibinfo {author} {\bibfnamefont {W.}~\bibnamefont
  {{Hikida}}},\ }\bibfield  {title} {\enquote {\bibinfo {title} {{Global
  structure of the Zipoy Voorhees Weyl spacetime and the $\delta$ = 2 Tomimatsu
  Sato spacetime}},}\ }\href {\doibase 10.1088/0264-9381/20/23/011} {\bibfield
  {journal} {\bibinfo  {journal} {Class. Quantum Grav.}\ }\textbf {\bibinfo
  {volume} {20}},\ \bibinfo {pages} {5121--5140} (\bibinfo {year} {2003})},\
  \Eprint {http://arxiv.org/abs/gr-qc/0304064} {gr-qc/0304064} \BibitemShut
  {NoStop}%
\bibitem [{\citenamefont {{Rubin}}\ \emph {et~al.}(1992)\citenamefont
  {{Rubin}}, \citenamefont {{Graham}},\ and\ \citenamefont {{Kenney}}}]{Rubin}%
  \BibitemOpen
  \bibfield  {author} {\bibinfo {author} {\bibfnamefont {V.~C.}\ \bibnamefont
  {{Rubin}}}, \bibinfo {author} {\bibfnamefont {J.~A.}\ \bibnamefont
  {{Graham}}}, \ and\ \bibinfo {author} {\bibfnamefont {J.~D.~P.}\ \bibnamefont
  {{Kenney}}},\ }\bibfield  {title} {\enquote {\bibinfo {title} {{Cospatial
  counterrotating stellar disks in the Virgo E7/S0 galaxy NGC 4550}},}\ }\href
  {\doibase 10.1086/186460} {\bibfield  {journal} {\bibinfo  {journal}
  {Astrophys. J. Lett.}\ }\textbf {\bibinfo {volume} {394}},\ \bibinfo {pages}
  {L9--L12} (\bibinfo {year} {1992})}\BibitemShut {NoStop}%
\bibitem [{\citenamefont {{Iodice}}\ and\ \citenamefont
  {{Corsini}}(2014)}]{Corsini}%
  \BibitemOpen
  \bibinfo {editor} {\bibfnamefont {E.}~\bibnamefont {{Iodice}}}\ and\ \bibinfo
  {editor} {\bibfnamefont {E.~M.}\ \bibnamefont {{Corsini}}},\ eds.,\
  \href@noop {} {\emph {\bibinfo {title} {Multi-Spin Galaxies}}},\ \bibinfo
  {series} {Astronomical Society of the Pacific Conference Series}, Vol.\
  \bibinfo {volume} {486}\ (\bibinfo {year} {2014})\BibitemShut {NoStop}%
\bibitem [{\citenamefont {{Bassett}}\ \emph {et~al.}(2017)\citenamefont
  {{Bassett}}, \citenamefont {{Bekki}}, \citenamefont {{Cortese}},\ and\
  \citenamefont {{Couch}}}]{Bass}%
  \BibitemOpen
  \bibfield  {author} {\bibinfo {author} {\bibfnamefont {R.}~\bibnamefont
  {{Bassett}}}, \bibinfo {author} {\bibfnamefont {K.}~\bibnamefont {{Bekki}}},
  \bibinfo {author} {\bibfnamefont {L.}~\bibnamefont {{Cortese}}}, \ and\
  \bibinfo {author} {\bibfnamefont {W.}~\bibnamefont {{Couch}}},\ }\bibfield
  {title} {\enquote {\bibinfo {title} {{The formation of S0 galaxies with
  counter-rotating neutral and molecular hydrogen}},}\ }\href {\doibase
  10.1093/mnras/stx958} {\bibfield  {journal} {\bibinfo  {journal} {Mon. Not.
  R. Astron. Soc.}\ }\textbf {\bibinfo {volume} {471}},\ \bibinfo {pages}
  {1892--1909} (\bibinfo {year} {2017})},\ \Eprint
  {http://arxiv.org/abs/1704.08434} {arXiv:1704.08434} \BibitemShut {NoStop}%
\bibitem [{\citenamefont {{Dyda}}\ \emph {et~al.}(2015)\citenamefont {{Dyda}},
  \citenamefont {{Lovelace}}, \citenamefont {{Ustyugova}}, \citenamefont
  {{Romanova}},\ and\ \citenamefont {{Koldoba}}}]{Dyda}%
  \BibitemOpen
  \bibfield  {author} {\bibinfo {author} {\bibfnamefont {S.}~\bibnamefont
  {{Dyda}}}, \bibinfo {author} {\bibfnamefont {R.~V.~E.}\ \bibnamefont
  {{Lovelace}}}, \bibinfo {author} {\bibfnamefont {G.~V.}\ \bibnamefont
  {{Ustyugova}}}, \bibinfo {author} {\bibfnamefont {M.~M.}\ \bibnamefont
  {{Romanova}}}, \ and\ \bibinfo {author} {\bibfnamefont {A.~V.}\ \bibnamefont
  {{Koldoba}}},\ }\bibfield  {title} {\enquote {\bibinfo {title}
  {{Counter-rotating accretion discs}},}\ }\href {\doibase
  10.1093/mnras/stu2131} {\bibfield  {journal} {\bibinfo  {journal} {Mon. Not.
  R. Astron. Soc.}\ }\textbf {\bibinfo {volume} {446}},\ \bibinfo {pages}
  {613--621} (\bibinfo {year} {2015})},\ \Eprint
  {http://arxiv.org/abs/1408.5626} {arXiv:1408.5626} \BibitemShut {NoStop}%
\bibitem [{\citenamefont {{King}}\ \emph {et~al.}(2008)\citenamefont {{King}},
  \citenamefont {{Pringle}},\ and\ \citenamefont {{Hofmann}}}]{KiPr}%
  \BibitemOpen
  \bibfield  {author} {\bibinfo {author} {\bibfnamefont {A.~R.}\ \bibnamefont
  {{King}}}, \bibinfo {author} {\bibfnamefont {J.~E.}\ \bibnamefont
  {{Pringle}}}, \ and\ \bibinfo {author} {\bibfnamefont {J.~A.}\ \bibnamefont
  {{Hofmann}}},\ }\bibfield  {title} {\enquote {\bibinfo {title} {{The
  evolution of black hole mass and spin in active galactic nuclei}},}\ }\href
  {\doibase 10.1111/j.1365-2966.2008.12943.x} {\bibfield  {journal} {\bibinfo
  {journal} {Mon. Not. R. Astron. Soc.}\ }\textbf {\bibinfo {volume} {385}},\
  \bibinfo {pages} {1621--1627} (\bibinfo {year} {2008})},\ \Eprint
  {http://arxiv.org/abs/0801.1564} {arXiv:0801.1564} \BibitemShut {NoStop}%
\bibitem [{\citenamefont {{Stephani}}\ \emph {et~al.}(2009)\citenamefont
  {{Stephani}}, \citenamefont {{Kramer}}, \citenamefont {{MacCallum}},
  \citenamefont {{Hoenselaers}},\ and\ \citenamefont {{Herlt}}}]{Stephani2009}%
  \BibitemOpen
  \bibfield  {author} {\bibinfo {author} {\bibfnamefont {H.}~\bibnamefont
  {{Stephani}}}, \bibinfo {author} {\bibfnamefont {D.}~\bibnamefont
  {{Kramer}}}, \bibinfo {author} {\bibfnamefont {M.}~\bibnamefont
  {{MacCallum}}}, \bibinfo {author} {\bibfnamefont {C.}~\bibnamefont
  {{Hoenselaers}}}, \ and\ \bibinfo {author} {\bibfnamefont {E.}~\bibnamefont
  {{Herlt}}},\ }\href@noop {} {\emph {\bibinfo {title} {Exact Solutions of
  Einstein's Field Equations}}}\ (\bibinfo  {publisher} {Cambridge University
  Press},\ \bibinfo {year} {2009})\BibitemShut {NoStop}%
\bibitem [{\citenamefont {{Israel}}(1966)}]{Israel66}%
  \BibitemOpen
  \bibfield  {author} {\bibinfo {author} {\bibfnamefont {W.}~\bibnamefont
  {{Israel}}},\ }\bibfield  {title} {\enquote {\bibinfo {title} {{Singular
  hypersurfaces and thin shells in general relativity}},}\ }\href {\doibase
  10.1007/BF02710419} {\bibfield  {journal} {\bibinfo  {journal} {Nuovo Cimento
  B Serie}\ }\textbf {\bibinfo {volume} {44}},\ \bibinfo {pages} {1--14}
  (\bibinfo {year} {1966})}\BibitemShut {NoStop}%
\bibitem [{\citenamefont {{Barrab{\`e}s}}\ and\ \citenamefont
  {{Israel}}(1991)}]{Barrab}%
  \BibitemOpen
  \bibfield  {author} {\bibinfo {author} {\bibfnamefont {C.}~\bibnamefont
  {{Barrab{\`e}s}}}\ and\ \bibinfo {author} {\bibfnamefont {W.}~\bibnamefont
  {{Israel}}},\ }\bibfield  {title} {\enquote {\bibinfo {title} {{Thin shells
  in general relativity and cosmology: The lightlike limit}},}\ }\href
  {\doibase 10.1103/PhysRevD.43.1129} {\bibfield  {journal} {\bibinfo
  {journal} {\prd}\ }\textbf {\bibinfo {volume} {43}},\ \bibinfo {pages}
  {1129--1142} (\bibinfo {year} {1991})}\BibitemShut {NoStop}%
\bibitem [{\citenamefont {{Kucha{\v r}}}(1968)}]{Kuchar68}%
  \BibitemOpen
  \bibfield  {author} {\bibinfo {author} {\bibfnamefont {K.}~\bibnamefont
  {{Kucha{\v r}}}},\ }\bibfield  {title} {\enquote {\bibinfo {title} {{Charged
  shells in general relativity and their gravitational collapse}},}\ }\href
  {\doibase 10.1007/BF01698208} {\bibfield  {journal} {\bibinfo  {journal}
  {Czech. J. Phys.}\ }\textbf {\bibinfo {volume} {18}},\ \bibinfo {pages}
  {435--463} (\bibinfo {year} {1968})}\BibitemShut {NoStop}%
\bibitem [{\citenamefont {{Wald}}(1984)}]{Wald84}%
  \BibitemOpen
  \bibfield  {author} {\bibinfo {author} {\bibfnamefont {R.~M.}\ \bibnamefont
  {{Wald}}},\ }\href@noop {} {\emph {\bibinfo {title} {{General relativity}}}}\
  (\bibinfo  {publisher} {University of Chicago Press},\ \bibinfo {year}
  {1984})\BibitemShut {NoStop}%
\bibitem [{\citenamefont {{Boyer}}\ and\ \citenamefont
  {{Lindquist}}(1967)}]{BoyLindq67}%
  \BibitemOpen
  \bibfield  {author} {\bibinfo {author} {\bibfnamefont {R.~H.}\ \bibnamefont
  {{Boyer}}}\ and\ \bibinfo {author} {\bibfnamefont {R.~W.}\ \bibnamefont
  {{Lindquist}}},\ }\bibfield  {title} {\enquote {\bibinfo {title} {{Maximal
  Analytic Extension of the Kerr Metric}},}\ }\href {\doibase
  10.1063/1.1705193} {\bibfield  {journal} {\bibinfo  {journal} {J. Math.
  Phys.}\ }\textbf {\bibinfo {volume} {8}},\ \bibinfo {pages} {265--281}
  (\bibinfo {year} {1967})}\BibitemShut {NoStop}%
\bibitem [{\citenamefont {{Hoenselaers}}(1979)}]{Hoenselaers1979}%
  \BibitemOpen
  \bibfield  {author} {\bibinfo {author} {\bibfnamefont {C.}~\bibnamefont
  {{Hoenselaers}}},\ }\bibfield  {title} {\enquote {\bibinfo {title} {{Weyl
  conform tensor of the Tomimatsu-Sato {$\delta$} = 3 metric}},}\ }\href
  {\doibase 10.1007/BF00759273} {\bibfield  {journal} {\bibinfo  {journal}
  {General Relativity and Gravitation}\ }\textbf {\bibinfo {volume} {11}},\
  \bibinfo {pages} {325--327} (\bibinfo {year} {1979})}\BibitemShut {NoStop}%
\bibitem [{\citenamefont {{Tomimatsu}}\ and\ \citenamefont
  {{Sato}}(1972)}]{TS72PRL}%
  \BibitemOpen
  \bibfield  {author} {\bibinfo {author} {\bibfnamefont {A.}~\bibnamefont
  {{Tomimatsu}}}\ and\ \bibinfo {author} {\bibfnamefont {H.}~\bibnamefont
  {{Sato}}},\ }\bibfield  {title} {\enquote {\bibinfo {title} {{New Exact
  Solution for the Gravitational Field of a Spinning Mass}},}\ }\href {\doibase
  10.1103/PhysRevLett.29.1344} {\bibfield  {journal} {\bibinfo  {journal}
  {Phys. Rev. Lett.}\ }\textbf {\bibinfo {volume} {29}},\ \bibinfo {pages}
  {1344--1345} (\bibinfo {year} {1972})}\BibitemShut {NoStop}%
\bibitem [{\citenamefont {{Carter}}(1968)}]{Carter1968}%
  \BibitemOpen
  \bibfield  {author} {\bibinfo {author} {\bibfnamefont {B.}~\bibnamefont
  {{Carter}}},\ }\bibfield  {title} {\enquote {\bibinfo {title} {{Global
  Structure of the Kerr Family of Gravitational Fields}},}\ }\href {\doibase
  10.1103/PhysRev.174.1559} {\bibfield  {journal} {\bibinfo  {journal} {Phy.
  Rev.}\ }\textbf {\bibinfo {volume} {174}},\ \bibinfo {pages} {1559--1571}
  (\bibinfo {year} {1968})}\BibitemShut {NoStop}%
\bibitem [{\citenamefont {{Tahvildar-Zadeh}}(2015)}]{ZadehJMP2015}%
  \BibitemOpen
  \bibfield  {author} {\bibinfo {author} {\bibfnamefont {A.~S.}\ \bibnamefont
  {{Tahvildar-Zadeh}}},\ }\bibfield  {title} {\enquote {\bibinfo {title} {{On a
  zero-gravity limit of the Kerr-Newman spacetimes and their electromagnetic
  fields}},}\ }\href {\doibase 10.1063/1.4915290} {\bibfield  {journal}
  {\bibinfo  {journal} {J. Math. Phys.}\ }\textbf {\bibinfo {volume} {56}},\
  \bibinfo {eid} {042501} (\bibinfo {year} {2015})}\BibitemShut {NoStop}%
\bibitem [{\citenamefont {{Brauer}}\ \emph {et~al.}(2015)\citenamefont
  {{Brauer}}, \citenamefont {{Camargo}},\ and\ \citenamefont
  {{Socolovsky}}}]{Brauer2015}%
  \BibitemOpen
  \bibfield  {author} {\bibinfo {author} {\bibfnamefont {O.}~\bibnamefont
  {{Brauer}}}, \bibinfo {author} {\bibfnamefont {H.~A.}\ \bibnamefont
  {{Camargo}}}, \ and\ \bibinfo {author} {\bibfnamefont {M.}~\bibnamefont
  {{Socolovsky}}},\ }\bibfield  {title} {\enquote {\bibinfo {title}
  {{Newman-Janis Algorithm Revisited}},}\ }\href {\doibase
  10.1007/s10773-014-2225-3} {\bibfield  {journal} {\bibinfo  {journal} {Int.
  J. Theor. Phys.}\ }\textbf {\bibinfo {volume} {54}},\ \bibinfo {pages}
  {302--314} (\bibinfo {year} {2015})},\ \Eprint
  {http://arxiv.org/abs/1404.1949} {arXiv:1404.1949 [gr-qc]} \BibitemShut
  {NoStop}%
\bibitem [{\citenamefont {{Bi\v{c}\'{a}k}}\ \emph {et~al.}(1993)\citenamefont
  {{Bi\v{c}\'{a}k}}, \citenamefont {{Semer\' ak}},\ and\ \citenamefont
  {{Hadrava}}}]{Semerak1993}%
  \BibitemOpen
  \bibfield  {author} {\bibinfo {author} {\bibfnamefont {J.}~\bibnamefont
  {{Bi\v{c}\'{a}k}}}, \bibinfo {author} {\bibfnamefont {O.}~\bibnamefont
  {{Semer\' ak}}}, \ and\ \bibinfo {author} {\bibfnamefont {P.}~\bibnamefont
  {{Hadrava}}},\ }\bibfield  {title} {\enquote {\bibinfo {title} {{Collimation
  Effects of the Kerr Field}},}\ }\href {\doibase 10.1093/mnras/263.3.545}
  {\bibfield  {journal} {\bibinfo  {journal} {Mon. Not. R. Astron. Soc.}\
  }\textbf {\bibinfo {volume} {263}},\ \bibinfo {pages} {545--559} (\bibinfo
  {year} {1993})}\BibitemShut {NoStop}%
\bibitem [{\citenamefont {{Neugebauer}}\ and\ \citenamefont
  {{Meinel}}(1993)}]{NM1993ApJ}%
  \BibitemOpen
  \bibfield  {author} {\bibinfo {author} {\bibfnamefont {G.}~\bibnamefont
  {{Neugebauer}}}\ and\ \bibinfo {author} {\bibfnamefont {R.}~\bibnamefont
  {{Meinel}}},\ }\bibfield  {title} {\enquote {\bibinfo {title} {{The
  Einsteinian gravitational field of the rigidly rotating disk of dust}},}\
  }\href {\doibase 10.1086/187005} {\bibfield  {journal} {\bibinfo  {journal}
  {Astrophys. J. Lett.}\ }\textbf {\bibinfo {volume} {414}},\ \bibinfo {pages}
  {L97--L99} (\bibinfo {year} {1993})}\BibitemShut {NoStop}%
\bibitem [{\citenamefont {{Perj{\'e}s}}(1989)}]{Perjes89}%
  \BibitemOpen
  \bibfield  {author} {\bibinfo {author} {\bibfnamefont {Z.}~\bibnamefont
  {{Perj{\'e}s}}},\ }\bibfield  {title} {\enquote {\bibinfo {title} {{Factor
  structure of the Tomimatsu-Sato metrics}},}\ }\href {\doibase
  10.1063/1.528545} {\bibfield  {journal} {\bibinfo  {journal} {J. Math.
  Phys.}\ }\textbf {\bibinfo {volume} {30}},\ \bibinfo {pages} {2197--2200}
  (\bibinfo {year} {1989})}\BibitemShut {NoStop}%
\end{thebibliography}%

\end{document}